\documentclass[10pt,letterpaper,twocolumn,aps,pra, superscriptaddress,longbibliography]{revtex4-1}
\pdfoutput=1

\usepackage{layouts}
\usepackage{physics}
\usepackage{ucs}
\usepackage{amsmath}
\usepackage{amsfonts}
\usepackage{amssymb}
\usepackage{makeidx}
\usepackage{cellspace,booktabs}
\usepackage{natbib}
\usepackage{hyperref}
\usepackage[capitalise]{cleveref} 
\usepackage{xcolor}
\usepackage{microtype}
\usepackage{bm}
\usepackage{graphicx}
\usepackage{multirow}

\usepackage{lipsum}

\renewcommand{\dag}{^{\dagger}}

\begin{document}
\date{\today}

\author{Christian~Kraglund~Andersen}
\email{christian.andersen@phys.ethz.ch}
\affiliation{Department of Physics, ETH Zurich, CH-8093 Zurich, Switzerland}
\author{Ants~Remm}
\affiliation{Department of Physics, ETH Zurich, CH-8093 Zurich, Switzerland}
\author{Stefania~Lazar}
\affiliation{Department of Physics, ETH Zurich, CH-8093 Zurich, Switzerland}
\author{Sebastian~Krinner}
\affiliation{Department of Physics, ETH Zurich, CH-8093 Zurich, Switzerland}
\author{Nathan~Lacroix}
\affiliation{Department of Physics, ETH Zurich, CH-8093 Zurich, Switzerland}
\author{Graham~J.~Norris}
\affiliation{Department of Physics, ETH Zurich, CH-8093 Zurich, Switzerland}
\author{Mihai~Gabureac}
\affiliation{Department of Physics, ETH Zurich, CH-8093 Zurich, Switzerland}
\author{Christopher~Eichler}
\affiliation{Department of Physics, ETH Zurich, CH-8093 Zurich, Switzerland}
\author{Andreas~Wallraff}
\affiliation{Department of Physics, ETH Zurich, CH-8093 Zurich, Switzerland}

\title{Repeated Quantum Error Detection in a Surface Code}

\date{\today}

\begin{abstract}
  The realization of quantum error correction is an essential ingredient for reaching the full potential of fault-tolerant universal quantum computation. Using a range of different schemes, logical qubits can be redundantly encoded in a set of physical qubits. One such scalable approach is based on the surface code.  Here we experimentally implement its smallest viable instance, capable of repeatedly detecting any single error using seven superconducting qubits, four data qubits and three ancilla qubits. Using high-fidelity ancilla-based stabilizer measurements we initialize the cardinal states of the encoded logical qubit with an average logical fidelity of 96.1\%. We then repeatedly check for errors using the stabilizer readout and observe that the logical quantum state is preserved with a lifetime and coherence time longer than those of any of the constituent qubits when no errors are detected. Our demonstration of error detection with its resulting enhancement of the conditioned logical qubit coherence times in a 7-qubit surface code is an important step indicating a promising route towards the realization of quantum error correction in the surface code.

\end{abstract}
\maketitle

\section*{Introduction}
The feasibility of quantum simulations and computations with more than 50 qubits has been demonstrated in recent experiments~\cite{Zhang2017k, Bernien2017, Arute2019}. Many near-term efforts in quantum computing are currently focused on the implementation of applications for noisy intermediate-scale quantum devices~\cite{Preskill2018}. However, to harness the full potential of quantum computers, fault tolerant quantum computing must be implemented. Quantum error correction and fault-tolerance have been explored experimentally in a variety of physical platforms such as nuclear magnetic resonance~\cite{Cory1998}, trapped ions~\cite{Chiaverini2004, Schindler2011, Lanyon2013, Linke2017a}, photonics~\cite{Yao2012a, Bell2014}, NV-centers~\cite{Cramer2016}, and superconducting circuits~\cite{Reed2012, Shankar2013a, Riste2015, Kelly2015, Corcoles2015, Ofek2016}. In particular, recent experiments have demonstrated quantum state stabilization~\cite{Riste2013, Negnevitsky2018, Andersen2019, Bultink2019}, simple error correction codes~\cite{Schindler2011, Riste2015, Kelly2015, Nigg2014b, Gong2019a} and the fault-tolerant encoding of logical quantum states~\cite{Linke2017a, Takita2017}. Quantum error correction with logical error rates comparable or below that of the physical constituents has also been achieved encoding quantum information in continuous variables using superconducting circuits~\cite{Ofek2016, Hu2019a, Campagne-Ibarcq2019}. These bosonic encoding schemes take advantage of high quality factor microwave cavities which are predominantly limited by photon loss. However, so far no repeated detection of both amplitude and phase errors on an encoded logical qubit has been realized in any qubit architecture. In this work, we present such a demonstration using the surface code~\cite{Fowler2012}, which, due to its high error-threshold, is one of the most promising architectures for large-scale fault-tolerant quantum computing.

\begin{figure*}[ht]
\includegraphics[width=\linewidth]{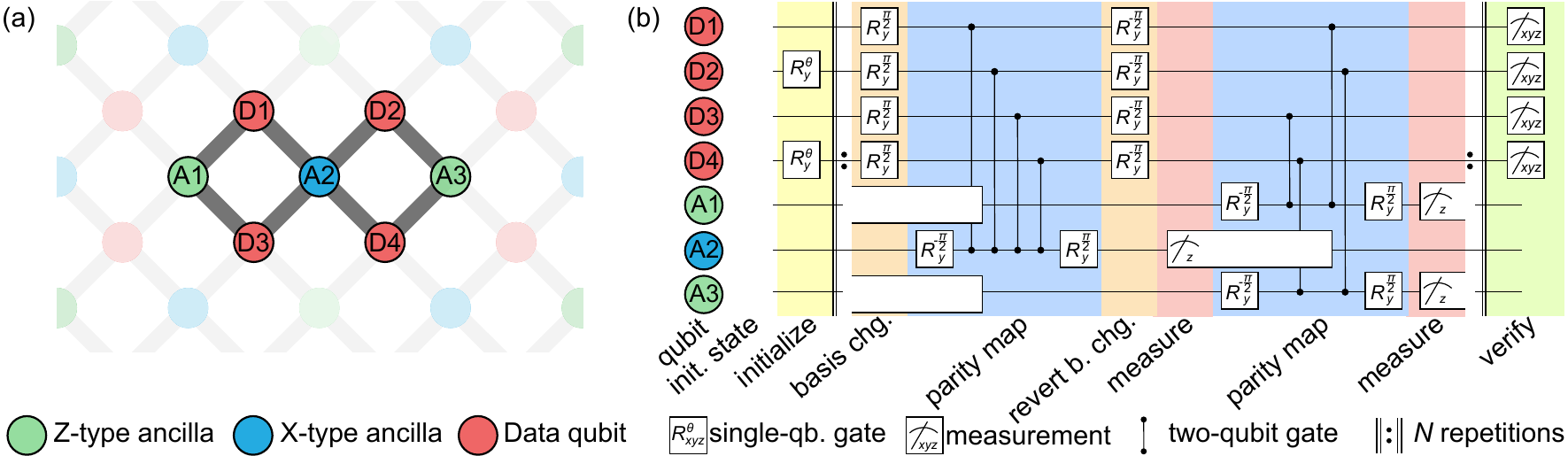}
\caption{Seven qubit surface code. (a) The surface code consists of a two-dimensional array of qubits. Here the data qubits are shown in red an the ancilla qubits for measuring $X$-type ($Z$-type) stabilizers in blue (green). The smallest surface code consists of seven qubits indicated by the data qubits D1-D4 and the ancilla qubits A1-A3. (b) Gate sequence for quantum error detection using the seven qubit surface code. Details of the gate sequence are discussed in the main text.} \label{fig:fig1}
\end{figure*}

In stabilizer codes for quantum error correction~\cite{Lidar2013, Terhal2015n}, a set of commuting multi-qubit operators is repeatedly measured, which projects the qubits onto a degenerate eigenspace of the stabilizers referred to as the code space. Thus, the experimental realization of quantum error detection crucially relies on high-fidelity entangling gates between the data qubits and the ancilla qubits and on the simultaneous high-fidelity single-shot readout of all ancilla qubits. For superconducting circuits, multiplexed readout has recently been implemented for high-fidelity simultaneous readout in multi-qubit architectures~\cite{Groen2013, Schmitt2014a, Jeffrey2014} with small crosstalk~\cite{Heinsoo2018}. Small readout crosstalk leading to minimal unwanted dephasing of data qubits when performing ancilla readout has been key enabler of recent experiments in superconducting circuits realizing repeated acilla-based parity detection~\cite{Andersen2019, Bultink2019}. Moreover, repeatable high-fidelity single- and two-qubit gates~\cite{Barends2014, Rol2019}, required for quantum error correction, have also been demonstrated for superconducting qubits.
Here, we utilize low-crosstalk multiplexed readout and a sequential stabilizer-measurement scheme~\cite{Versluis2017} for implementing a seven qubit surface code with superconducting circuits.

In the surface code, as in any stabilizer code, errors are detected by observing changes in the stabilizer measurement outcomes. Such syndromes are typically measured by entangling the stabilizer operators with the state of ancilla qubits, which are then projectively measured to yield the stabilizer outcomes.
The surface code consists of a $d\times d$ grid of data qubits with $d^2{-}1$ ancilla qubits, each connected to up to four data qubits~\cite{Fowler2012}. The code can detect $d-1$ errors and correct up to $\lfloor(d-1)/2\rfloor$ errors per cycle of stabilizer measurements. In particular, the stabilizers of the $d=2$ surface code, see Fig.~\ref{fig:fig1}, are given by
\begin{align}
X_{D1} X_{D2} X_{D3} X_{D4},\qquad Z_{D1} Z_{D3}, \qquad Z_{D2} Z_{D4}.
\end{align}
For the code-distance $d=2$, it is only possible to detect a single error per round of stabilizer measurements and once an error is detected, the error can not be unambiguously identified, e.g. one would obtain the same syndrome outcome for an $X$-error on D1 and on D3.

Here, we use the following logical qubit operators
\begin{align}
Z_L = Z_{D1} Z_{D2}, \quad\text{or}\quad Z_L = Z_{D3} Z_{D4}, \label{eq:zl}\\
X_L = X_{D1} X_{D3}, \quad\text{or}\quad X_L = X_{D2} X_{D4}, \label{eq:xl}
\end{align}
such that the code space in terms of the physical qubit states is spanned by the logical qubit states
\begin{align}
\ket{0}_L = \frac{1}{\sqrt{2}}(\ket{0000} + \ket{1111}), \\
\ket{1}_L = \frac{1}{\sqrt{2}}(\ket{0101} + \ket{1010}).
\end{align}
To encode quantum information in the logical subspace, we initialize the data qubits in a separable state, chosen such that after a single cycle of stabilizer measurements and conditioned on ancilla measurement outcomes being $\ket{0}$, the data qubits are encoded into the target logical qubit state.
In this work, we demonstrate this probabilistic preparation scheme for the logical states $\ket{0}_L$, $\ket{1}_L$, $\ket{+}_L = (\ket{0}_L + \ket{1}_L)/\sqrt{2}$ and $\ket{-}_L = (\ket{0}_L - \ket{1}_L)/\sqrt{2}$ and we perform repeated error detection on these states.

\begin{figure*}
\includegraphics[width=\linewidth]{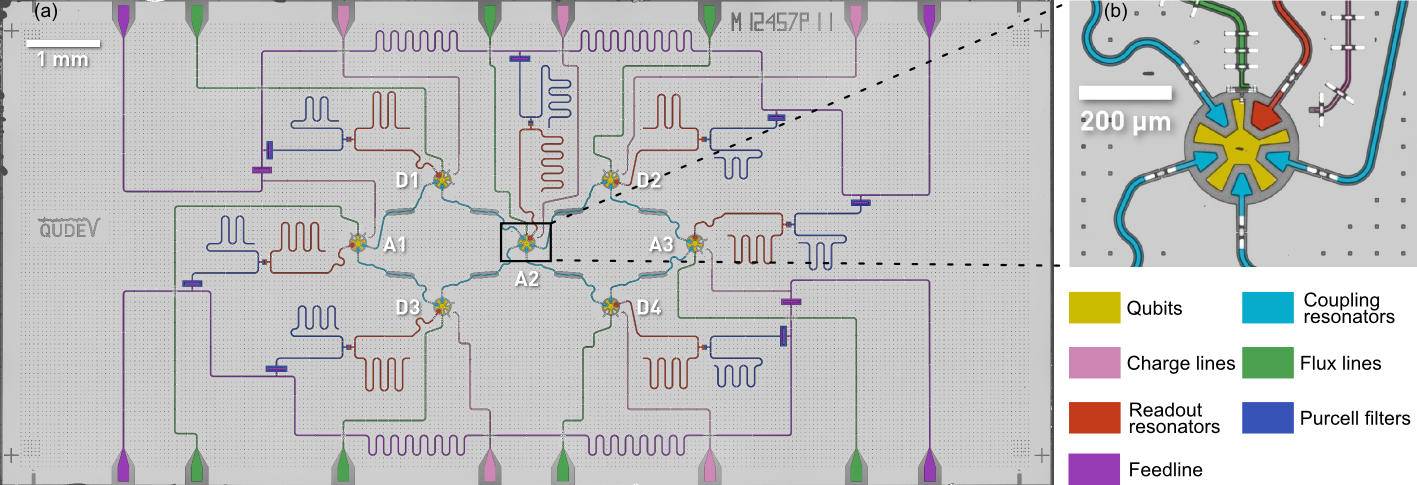}
\caption{Seven-qubit device. (a) False colored micrograph of the seven-qubit device used in this work. Transmon qubits are shown in yellow, coupling resonators in cyan, flux lines for single-qubit tuning and two-qubit gates in green, charge lines for single-qubit drive in pink, the two feedlines for readout in purple, transmission line resonators for readout in red and Purcell filters for each qubit in blue. (b) Enlarged view of the center qubit (A2) which connects to four neighboring qubits.} \label{fig:chip}
\end{figure*}

\section*{Implementation}
The seven qubit surface code, as discussed above, can be realized with a set of qubits laid out as depicted in Fig.~\ref{fig:fig1}(a). The logical qubit is encoded into four data qubits, D1-D4, and three ancilla qubits, A1, A2 and A3 are used to measure the three stabilizers $Z_{D1} Z_{D3}$, $X_{D1} X_{D2} X_{D3} X_{D4}$ and $Z_{D2} Z_{D4}$, respectively. We initially herald all qubits in the $\ket{0}$-state~\cite{Johnson2012, Riste2012} and subsequently prepare the data qubits in a product state using single qubit rotations around the $y$-axis. These initial states are then projected onto the code space after the initial stabilizer measurement cycles.

We perform the $X_{D1} X_{D2} X_{D3} X_{D4}$ stabilizer measurement by first applying basis change pulses ($R_Y^{\pi/2}$) on the data qubits to map the $X$ basis to the $Z$ basis. Then we perform the entangling gates as in Fig.~\ref{fig:fig1}(b) and finally we revert the basis change. The measurement of A2 will therefore yield the $\ket{0}$-state ($\ket{1}$-state) corresponding to the eigenvalue $+1$ ($-1$) of the stabilizer $X_{D1} X_{D2} X_{D3} X_{D4}$. While the measurement pulse for A2 is still being applied, we perform the $Z_{D1} Z_{D3}$ and $Z_{D2} Z_{D4}$  stabilizer measurements simultaneously using the ancilla qubits A1 and A3, respectively. To avoid unwanted interactions during entangling gate operations, we operate the surface code using a pipelined approach similar to the scheme introduced by Versluis et.al.~\cite{Versluis2017}, for which we perform $X$-type and $Z$-type stabilizer measurements sequentially, see Fig.~\ref{fig:fig1}(b) and Appendix~\ref{app:pulse_seq}. The cycle is repeated after this step, and, after $N$ stabilizer measurement cycles, we perform state tomography of the data qubits.

The gate sequence described above is implemented on the seven qubit superconducting quantum device shown in Fig.~\ref{fig:chip}(a), see Appendix~\ref{app:device} for device parameters. Each qubit (yellow) is a single-island transmon qubit~\cite{Koch2009} and features an individual flux line (green) for frequency tuning and an individual charge line (pink) for single qubit gates. Additionally, each qubit is coupled to a readout resonator (red) combined with an individual Purcell filter (blue). The Purcell filters protect against qubit decay into the readout circuit~\cite{Reed2010} and suppress readout crosstalk such that multiplexed ancilla measurements can be performed without detrimental effects on the data qubits~\cite{Heinsoo2018}. Each Purcell filter is coupled to a feedline and we perform all measurements by probing each feedline with a frequency-multiplexed readout pulse~\cite{Heinsoo2018}, see Appendix~\ref{app:readout} for a complete characterization of the readout. The qubits are coupled to each other via 1.5~mm long coplanar waveguide segments (cyan) as displayed in Fig.~\ref{fig:fig1}(a). The seven qubit surface code requires the central ancilla  qubit to connect to four neighbors. The qubit island shape, shown Fig.~\ref{fig:chip}(b), is designed to facilitate coupling to a readout resonator and four two-qubit couplers. To ensure a closed ground plane around the qubit island, each coupler element crosses the ground plane with an airbridge (white). We install the device in a cryogenic measurement setup~\cite{Krinner2019}, see Appendix~\ref{app:setup}, and we characterize and benchmark the device using time-domain and randomized benchmarking methods as detailed in Appendix~\ref{app:device}.

\section*{Results}

\begin{figure}[b]
\includegraphics[width=\linewidth]{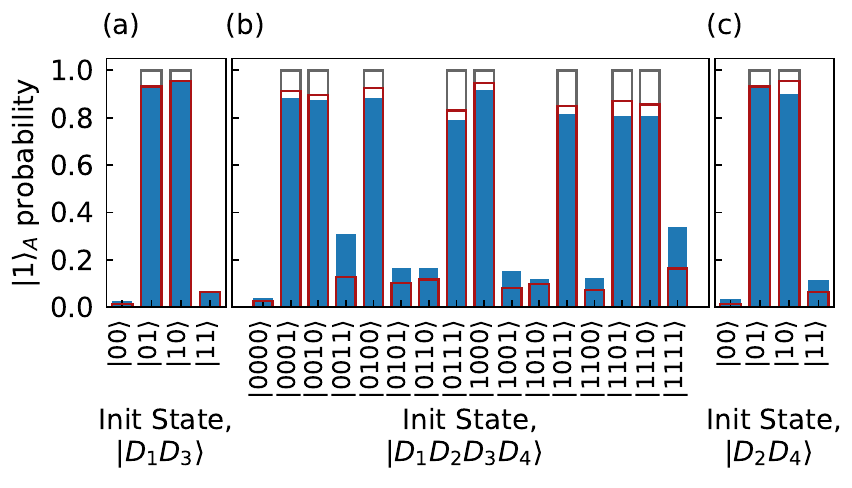}
\caption{Stabilizer measurements of the data qubits. In (a) we show the outcomes of the measurement of $Z_{D1}Z_{D3}$ using ancilla A=A1, in (b) of $Z_{D1} Z_{D2} Z_{D3} Z_{D4}$ using ancilla A=A2 and in (c) of $Z_{D2}Z_{D4}$ using ancilla A=A3. For all panels we show the ideal outcome in the gray wireframe and the corresponding master equation simulations in the red wireframe.
} \label{fig:parity}
\end{figure}

Changes in the outcome of repeated stabilizer measurements, also referred to as syndromes, signal the occurrence of an error. It is, thus, critical to directly verify the ability to measure the multi-qubit stabilizers using the ancilla readout~\cite{Takita2016}. We characterize the performance of the stabilizer measurements by preparing the data qubits in each of the  computational basis states and measure the $Z$-stabilizers, see Fig.~\ref{fig:parity}. For each stabilizer, the other ancilla qubits and unused data qubits are left in the ground state.
We correctly assign the ancilla measurement outcome corresponding to the prepared basis state with success probabilities of $95.0\%$, $83.5\%$ and $91.8\%$ for the stabilizers $Z_{D1} Z_{D3} $, $Z_{D1} Z_{D2} Z_{D3} Z_{D4}$ and $Z_{D2} Z_{D4}$ calculated as the overlap between the measured probabilities and the ideal case (gray wireframe in Fig.~\ref{fig:parity}). Master equation simulations, which include decoherence and readout errors, are shown by the red wireframes in Fig.~\ref{fig:parity}. The parity measurements are mainly limited by the relaxation of the data qubits, which directly leads to worse results for states with multiple excitations such as the $\ket{1111}$-state when measuring $Z_{D1} Z_{D2} Z_{D3} Z_{D4}$. Further variations in the correct parity assignment probability arise due to the differences in qubit lifetimes and two-qubit gate durations (see Appendix~\ref{app:device}).

In a next step, we prepare logical states by projecting the data qubits onto the desired code space. We use a probabilistic encoding scheme, where we initialize the data qubits in a given product state and perform one cycle of stabilizer measurements. Then, in the events where all syndrome results are $\ket{0}$, the data qubits are projected onto the desired logical state. We can use this probabilistic scheme to prepare any logical state by initializing the state $\ket{0}(a\ket{0}+b\ket{1})\ket{0}(a\ket{0}+be^{i\phi}\ket{1})$, which will be projected onto the (unnormalized) logical state $\ket{\psi}_L = a^2 \ket{0}_L + b^2 e^{i\phi} \ket{1}_L$. Here, we specifically initialize the logical states $\ket{0}_L$, $\ket{1}_L$, $\ket{+}_L$ and $\ket{-}_L$  by performing one cycle of stabilizer measurements on the states $\ket{0000}$, $\ket{0101}$, $\ket{0{+}0{+}}$ and $\ket{0{+}0{-}}$, respectively, with $\ket{\pm} = (\ket{0} \pm \ket{1})/\sqrt{2}$.

\begin{figure}[b]
\includegraphics[width=\linewidth]{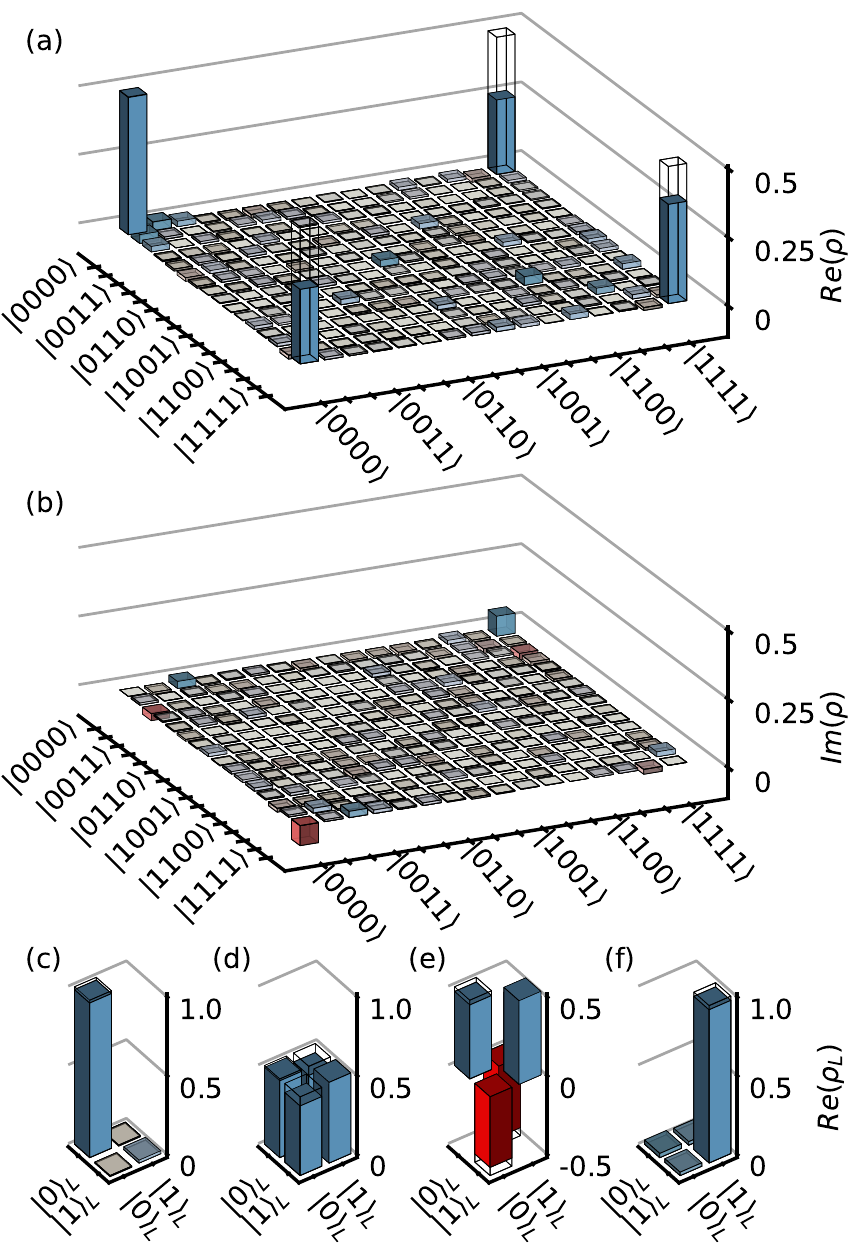}
\caption{Preparation of logical states. (a) Real and (b) imaginary part of the density matrix of the four physical data qubits prepared in the $\ket{0}_L$-state using a single round of stabilizer measurements. The fidelity to the target state, shown in the wire frame, is $F_{\mathrm{phys}}=70.3\%$. (c) Real part of $\rho_L$, i.e. the density matrix shown in (a) and (b), projected onto the logical subspace. The fidelity to the target logical state is $F_L=98.2$\%. (d), (e) and (f) Density matrices for the logical states $\ket{+}_L$, $\ket{-}_L$ and $\ket{1}_L$ respectively. The corresponding logical fidelities, $F_L$, are 94.2\%, 94.8\% and 97.3\%. } 
\label{fig:density}
\end{figure}

First, we consider the preparation of $\ket{0}_L$ for which the data qubit state $\ket{0000}$ after one cycle of stabilizer measurements is projected onto the state $\ket{\psi_{0}} = (\ket{0000} + \ket{1111})/\sqrt{2}$ when all ancilla qubits are measured in $\ket{0}$. We measure all ancilla qubits to be in the $\ket{0}$ state with a success-probability of 25.1\%, compared to an expected probability of 50\% in the ideal case. To verify the state preparation, we perform full state tomography of the four data qubits after the completion of one cycle of stabilizer measurements and construct the density matrix based on a maximum likelihood estimation taking readout errors into account. The measured density matrix of the physical data qubits has a fidelity of $F_{\mathrm{phys}}=\bra{\psi_{0}} \rho \ket{\psi_{0}} = 70.3\%$ to the target state, see Fig.~\ref{fig:density}(a-b). While the infidelity is dominated by qubit decoherence, we also observe small residual coherent phase errors as seen by the finite imaginary matrix elements in Fig.~\ref{fig:density}(b) corresponding to a phase error of 5 degrees accumulated over the cycle time of 1.92~$\mu$s or, equivalently, a frequency drift of 7~kHz for any qubit.

Given access to the full density matrix, we can project it onto the logical qubit subspace $\rho_{L,ji} = \bra{j}\rho\ket{i} / P_L$ for $\ket{i },\ket{j} \in \lbrace \ket{0}_L, \ket{1}_L \rbrace$. Here. $P_L = \sum_{i} \bra{i}\rho\ket{i}$ is the probability of the prepared state to be within the logical subspace, which is also referred to as~the acceptance probability~\cite{Takita2017} or yield~\cite{Linke2017a}. The state $\rho_L$ is the logical qubit state, conditioned on the prepared state residing in the code space at the end of the cycle. In general, the physical fidelity of the data qubits can be expressed in the form $F_{\mathrm{phys}} = F_L P_L$, where $F_L$ is the fidelity of $\rho_L$ compared to the ideal logical state. We experimentally find the probability $P_L = 0.717$ of the prepared state to be within the logical subspace. From simulations, we understand that the reduced $P_L$ mainly arises from decoherence during the stabilizer measurement cycle. After the projection onto the code space, the logical qubit state $\ket{0}_L$ is described by a single qubit density matrix, see Fig.~\ref{fig:density}(c), which has a fidelity of $F_L = 98.2\%$ to the ideal logical state. Similarly, we prepare the logical states $\ket{+}_L$, $\ket{-}_L$ and $\ket{1}_L$, shown in Fig.~\ref{fig:density}(d-f), with logical state fidelities of 94.2\%, 94.8\% and 97.3\%, respectively.
The corresponding logical fidelities of the four logical states from master equation simulations are 98.5\%, 96.6\%, 96.4\% and 98.1\%, see Appendix~\ref{app:sim}. The slightly lower fidelities for the $\ket{+}_L$ and $\ket{-}_L$ states arise from the pure dephasing of the data qubits making $Z$-errors during the encoding more likely than $X$-errors.

\begin{figure}
\includegraphics[width=\linewidth]{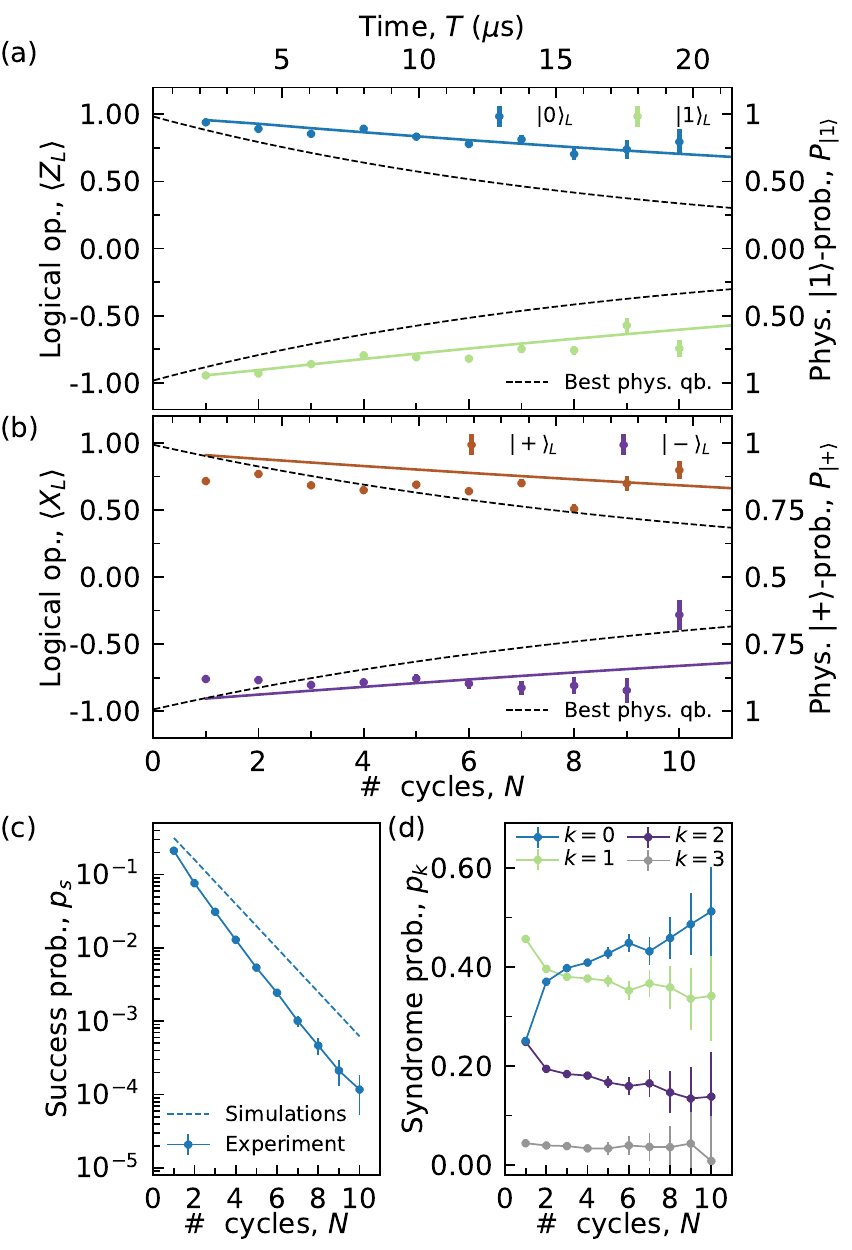}
\caption{Repeated quantum error detection. The expectation values of (a) the logical $Z_L$ operator and (b) the logical $X_L$ operator as a function of $N$, the number of stabilizer measurement cycles. The expectations values are shown for the prepared $\ket{0}_L$ (blue), $\ket{1}_L$ (green), $\ket{+}_L$ (brown) and $\ket{-}_L$ (purple) states. The solid lines indicate the corresponding values obtained from master equation simulations. Also shown (dashed lines, right axis) are the (a) qubit decay of the $\ket{1}$-state with the best measured $T_1$ value and (b) the physical qubit decay of the $\ket{+}$-state with the best measured $T_2$ value. (c) Total success probability $p_s$ for detecting no errors during $N$ cycles of stabilizer measurements for the $\ket{0}_L$ data shown in~(a) and the corresponding values from numerical simulations. (d) Probability of observing $k$ ancilla qubits in the $\ket{1}$ state for each measurement cycle and conditioned on having detected no error in any of the previous $N{-}1$ cycles. The data corresponds to the initial $\ket{0}_L$ state presented in~(a). } \label{fig:logical}
\end{figure}

Next, we demonstrate repeated quantum error detection of any single error, which is a key ingredient of quantum error correction schemes such as the surface code. We do so by repeatedly measuring the expectation value of the encoded qubit's logical $Z_L$ ($X_L$) operator conditioned on having detected no error in any repetition of the stabilizer measurement and on having the final measurement of the data qubits satisfy $Z_{D1}Z_{D3}=Z_{D2}Z_{D4}=1$ ($X_{D1} X_{D2} X_{D3} X_{D4} = 1$). This latter condition ensures that the qubits have remained in the logical subspace during the last detection cycle. We find that the expectation value $\langle Z_L \rangle$ (green and blue data points) decays in good approximation exponentially from unity with a logical life time of $62.7\pm9.4$~$\mu$s from this exponential fit, which exceeds the life time, 16.8~$\mu$s, of the best physical qubit (dashed lines) of the device, see Fig.~\ref{fig:logical}(a). The logical expectation values are evaluated after the $N$th cycle at time $T=(1.92N+0.3)\,\mu$s shown at the top axis of Fig.~\ref{fig:logical}(a,b). The approximately exponential decay of the logical qubit expectation value $\langle X_L \rangle$ (brown and purple points) indicates a logical coherence time $72.5\pm32.9$~$\mu$s, also exceeding that of the best physical qubit, 21.5~$\mu$s, on the device (dashed lines), Fig.~\ref{fig:logical}(b). However, the fits to $\langle X_L \rangle$ show larger error bars due to the finite fidelity of preparing the logical $\ket{+}_L$ and $\ket{-}_L$ states, limited by the pure dephasing of the qubits as also seen in Fig.~\ref{fig:density}(d-e). Converting the measured decay times into an error per stabilizer measurement cycle, we find a logical $X_L$ error probability of $3.1\%\pm0.45\%$ and a logical $Z_L$ error probability of $2.6\pm 1.3\%$.

Generally, we find good agreement between the measured expectation values of the logical qubit operators and the ones calculated using numerical simulations, solid lines in Fig.~\ref{fig:logical}(a,b), accounting for finite physical qubit life- ($T_1$) and coherence times ($T_2$), residual-$ZZ$ coupling and readout errors, see Appendix~\ref{app:sim} for details. From the numerical simulations, we extract logical decay times of $44.2$~$\mu$s and $59.6$~$\mu$s for $Z_L$ and $X_L$ operators when no errors are detected, which are smaller than the experimentally obtained times, but within the experimental error bars. The simulated decay times correspond to a logical $X_L$ error probability of 4.2\% and a logical $Z_L$ error probability of 3.1\% per error detection cycle. We suspect that for the $\ket{+}_L$-state coherent errors from qubit frequency drifts during the data collection cause the deviations between data and simulations.

Finally, we discuss the probability to observe $k$ ancilla qubits simultaneously in the $\ket{1}$ state per error detection cycle when no errors were detected in previous cycles.
We find that the probability to observe no errors slowly increases with $N$ from about 40\% to 50\%, see Fig.~\ref{fig:logical}(d). From numerical simulations, we find that the probability to observe no additional errors after one cycle is between 49.9\% and 50.3\% per cycle, slightly larger than the experimentally observed values. We also observe experimentally that the probability of detecting more than a single ancilla qubit in the $\ket{1}$ state per cycle is approximately suppressed exponentially. Consistent with this analysis, we find that the measured probability of not detecting an error (blue data points) decreases exponentially with $N$, Fig.~\ref{fig:logical}(c). After $N=10$ cycles, the success probability, i.e.~the total probability that the state remained in the code space, approaches~$10^{-4}$, around a factor of 6 smaller than the simulated value. The difference between the simulated (dashed line) and experimentally determined success probabilities stems from the smaller simulated error probability per cycle discussed above.

\section*{Discussion}
In conclusion, we have implemented a seven qubit surface code for repeated quantum error detection. In particular, our experiment was enabled by fast and low-crosstalk readout for ancilla measurements. Using the seven qubit surface code, we demonstrated preparation of the logical states $\ket{0}_L$, $\ket{1}_L$, $\ket{+}_L$ and $\ket{-}_L$ with an average fidelity in the logical subspace of 96.1\%. The probability to be within the logical subspace was found to be around 70\% due to the accumulated errors during the stabilizer measurement cycle in good agreement with the corresponding numerical simulations. When executing the quantum error detection sequence for multiple cycles, we find an extended lifetime and coherence time of the logical qubit conditioned on detecting no errors. The data presented here is postselected on the ancilla measurement outcomes and on the condition that the final measurement of the data qubits satisfies the stabilizer conditions of the code. Crucially, since we found both extended logical life- and coherence time, we verified that neither the syndrome measurements nor the postselection extract information about the logical quantum state. The techniques used in this work for high-fidelity gates~\cite{Rol2019} and low-crosstalk qubit readout~\cite{Heinsoo2018} are directly applicable to a range of error correction codes~\cite{Bacon2006, Bombin2006, Chamberland2019, Li2019s} which also critically require repeated measurements of ancilla qubits with minimal detrimental effects on the data qubits.
Our implementation uses a gate sequence that is extensible to large surface codes~\cite{Versluis2017} and, thus, our work represents a key demonstration towards using superconducting quantum devices for fault-tolerant quantum computing.


\section*{Data availability statement}
The data produced in this work is available from the corresponding author upon reasonable request.

\section*{Acknowledgments}
The authors are grateful for valuable feedback from K.~Brown and A.~Darmawan. The authors acknowledge contributions to the measurement setup from S.~Storz, F.~Swiadek and T.~Zellweger.

The authors acknowledge financial support by the Office of the Director of National Intelligence (ODNI), Intelligence Advanced Research Projects Activity (IARPA), via the U.S. Army Research Office grant W911NF-16-1-0071, by the National Centre of Competence in Research Quantum Science and Technology (NCCR QSIT), a research instrument of the Swiss National Science Foundation (SNSF), by the EU Flagship on Quantum Technology  H2020-FETFLAG-2018-03 project 820363 OpenSuperQ, by the SNFS R'equip grant 206021-170731 and by ETH Zurich. The views and conclusions contained herein are those of the authors and should not be interpreted as necessarily representing the official policies or endorsements, either expressed or implied, of the ODNI, IARPA, or the U.S. Government.

\section*{Author Contributions}
C.K.A. designed the device and A.R., S.K., G.N. and M.G fabricated the device. C.K.A., A.R., S.L. and N.L. developed the experimental control software. C.K.A., A.R., S.K. and N.L. installed the experimental setup. C.K.A., A.R. and S.L. characterized and calibrated the device and the experimental setup. C.K.A. carried out the main experiment and analyzed the data. C.K.A. performed the numerical simulations. C.E. and A.W. supervised the work. C.K.A., A.R. and S.L. prepared the figures for the manuscript. C.K.A. wrote the manuscript with input from all co-authors.

\section*{Competing interests}
The authors declare no competing interests.

\begin{appendix}
\section*{Supplementary Information}

\section{Pulse sequence}
\label{app:pulse_seq}

\begin{figure*}
\includegraphics[width=\linewidth]{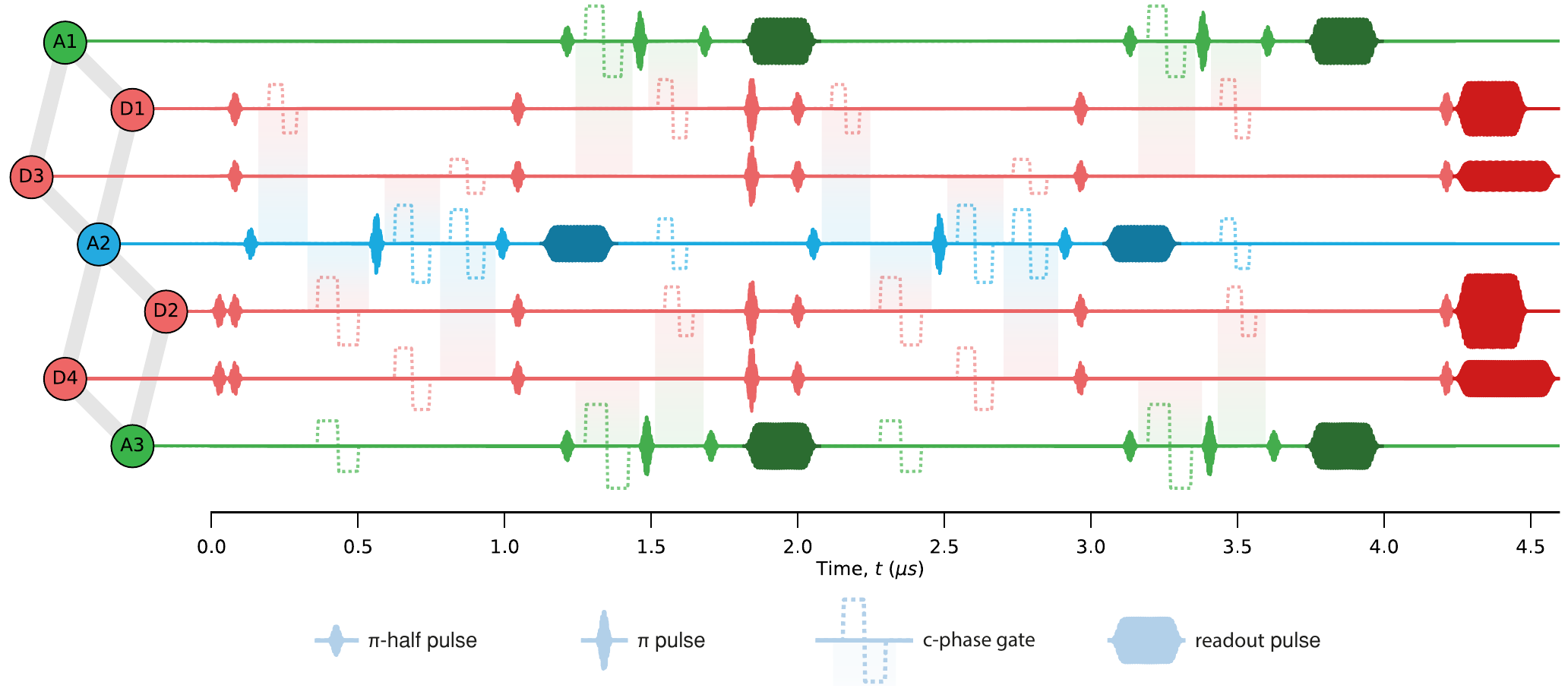}
\caption{AWG waveforms for two cycles of the stabilizer measurement. Solid lines represent the microwave pulses for single qubit gates, dark solid pulses the readout pulses, and dashed lines zero-area flux pulses. The shaded areas indicate which two qubits interact during each flux pulse. } \label{fig:ps}
\end{figure*}

We physically implement the gate sequence shown in Fig.~\ref{fig:fig1} by waveforms on the arbitrary waveform generators (AWGs) of the experimental setup, see Fig.~\ref{fig:ps} for an example with two cycles of stabilizer measurements. The pulse sequence includes dynamical decoupling pulses on the ancilla qubits during the stabilizer measurements and dynamical decoupling pulses on the data qubits in between each stabilizer cycle. All qubits are parked at their upper sweetspot which enable us to use the net-zero flux pulse shape as introduced by Rol et.al.~\cite{Rol2019}. The net-zero pulse is shaped such that the integral of the pulse is zero which serves to limit memory effects on the two-qubit gates e.g. due to charge accumulation in the flux lines. Beyond the flux pulses that enable the two-qubit gates (indicated with the shaded background), we apply additional flux pulses to non-interacting qubits. These additional flux pulses serves the purpose of pushing the frequency of the qubit down in frequency such that we avoid frequency collisions during the gate.

\begin{table*}[t]
\centering
\begin{tabular}{lccccccc}
\hline
\noalign{\vskip 1mm}
 & D1 & D2 & D3  & D4 & A1 & A2 & A3\\
 \hline
 \hline
Qubit frequency, $\omega_q/2\pi$ (GHz) & 5.494 & 5.712 & 4.108 & 4.222 & 4.852 & 4.963 & 5.190 \\
Lifetime, $T_1$ ($\mu$s) & 11.2 & 8.7 & 8.7 & 16.3 & 5.7 & 16.8 & 11.8 \\
Ramsey decay time, $T_2^*$ ($\mu$s) & 18.2 & 14.4 & 4.3 & 21.5 & 8.5 & 16.7 & 9.9 \\
Readout frequency, $\omega_r/2\pi$ (GHz) & 6.611 & 6.838 & 5.832 & 6.063 & 6.255 & 6.042 & 6.299\\
Readout linewidth, $\kappa_{\mathrm{eff}}/2\pi$ (MHz) & 7.5 & 10.6 & 6.0 & 7.2 & 17.3 & 10.9 & 11.0 \\
Purcell filter linewidth, $\kappa_{P}/2\pi$ (MHz) & 47.6 & 46.4 & 13.6 & 49.2 & 56.3 & 68.1 & 46.4 \\
Purcell-readout coupling, $J_{PR}/2\pi$ (MHz) & 20.0 & 22.2 & 17.5 & 18.4 & 18.8 & 18.7 & 19.0 \\
Purcell-readout detuning, $\Delta_{PR}/2\pi$ (MHz) & 33.8 & 25.7 & 19.4 & 32.3 & 11.3 & 25.6 & 20.6 \\
Dispersive shift, $\chi/2\pi$ (MHz) & -2.5 & -2.5 & -0.75 & -1.0 & -1.25 & -2.4 & -2.0 \\
Thermal population, $P_{\mathrm{th}}$ ($\%$) & 0.06 & 0.04 & 0.8 & 0.8 & 0.08 & 0.4 & 0.6 \\
Individual readout assignment prob. (\%) & 99.4 & 99.2 & 97.8 & 98.2 & 98.7 & 98.8 & 98.8 \\
Multiplexed readout assignment prob. (\%) & 98.9 & 99.1 & 98.2 & 97.4 & 97.7 & 98.4 & 98.6 \\
Measurement efficiency, $\eta$ & 0.30 & 0.24 & 0.15 & 0.15 & 0.20 & 0.27 & 0.22 \\
\hline
\end{tabular}
\caption{Measured parameters of the seven qubits.}
\label{tab:qb_params}
\end{table*}

\section{Device Fabrication and Characterization}
\label{app:device}

The device in Fig.~\ref{fig:chip} consists of seven qubits coupled to each other in the geometry showed in Fig.~\ref{fig:fig1}(a).
The resonator, coupling and qubit structures are defined using photolithography and reactive ion etching from a 150$\,$nm thin niobium film sputtered onto a high-resistivity intrinsic silicon substrate. To establish a well-connected ground plane, we add airbridges to the device. Airbridges are also used to cross signal lines, i.e., for the flux and charge lines to cross the feedlines. The aluminum-based Josephson junctions of the qubits are fabricated using electron beam lithography.

\begin{figure}
\includegraphics[width=\linewidth]{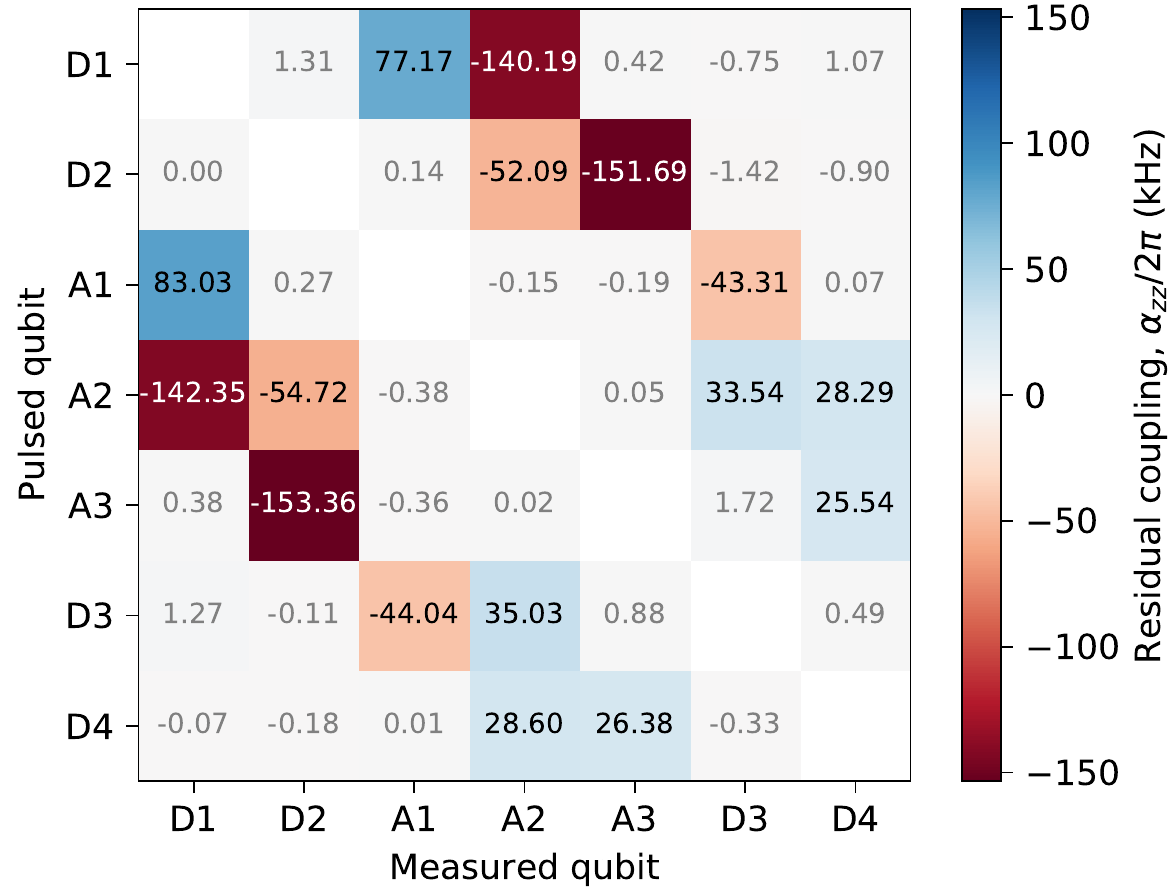}
\caption{Residual $ZZ$-coupling measured between all pairs of qubits on our device. The pulsed qubit is prepared in either ground or excited state and a Ramsey experiment is performed on the measured qubit to extract its frequency. The qubit pairs with gray label indicate pairs with no direct coupling.} \label{fig:resid_zz}
\end{figure}

\begin{figure}
\includegraphics[width=\linewidth]{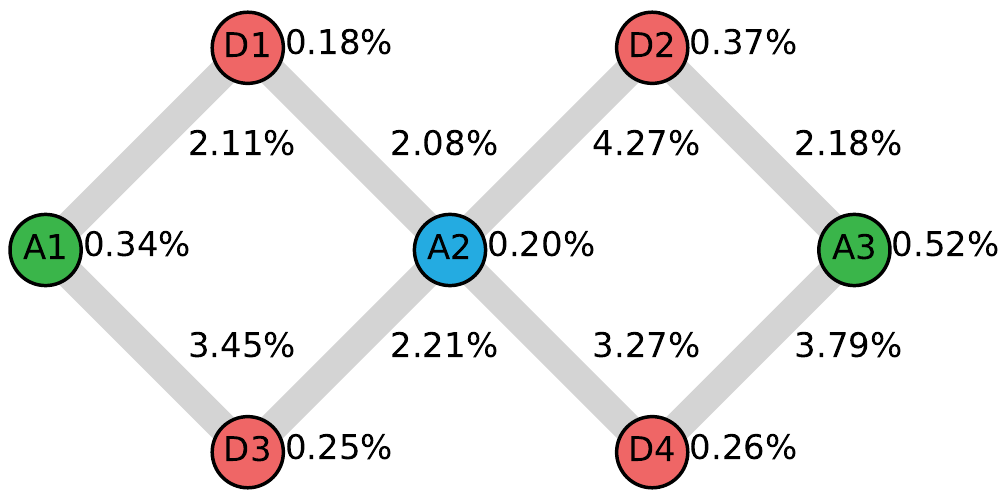}
\caption{Single qubit errors per Clifford for each qubit (circles), and average CZ gate errors from interleaved randomized benchamrking (lines).} \label{fig:rb}
\end{figure}

We extract the qubit parameters, see Table~\ref{tab:qb_params}, using standard spectroscopy and time domain methods. In addition to the parameters characterizing individual qubits, we measure the residual $ZZ$-coupling between all qubit pairs, see results in Fig.~\ref{fig:resid_zz}, by performing a Ramsey experiment on the measured qubit with the pulsed qubit in either the $\ket{0}$ or $\ket{1}$ state. To characterize the gate performance, we implement randomized benchmarking on all qubits to find the error per single qubit Clifford and we perform interleaved randomized benchmarking for the characterization of errors per conditional-phase gate. The resulting gate errors are shown in Fig.~\ref{fig:rb}. By directly measuring the $\ket{2}$-state population after the randomized benchmarking sequences, we further extract leakage per gate~\cite{Wood2017, Chen2016}. For single qubit gates, we find a leakage per Clifford operation to be $0.025\%$ on average while the leakage per conditional-phase gate is between $0.1\%$ and $0.7\%$.

\section{Readout Characterization}
\label{app:readout}

\begin{figure*}[t]
\includegraphics[width=\linewidth]{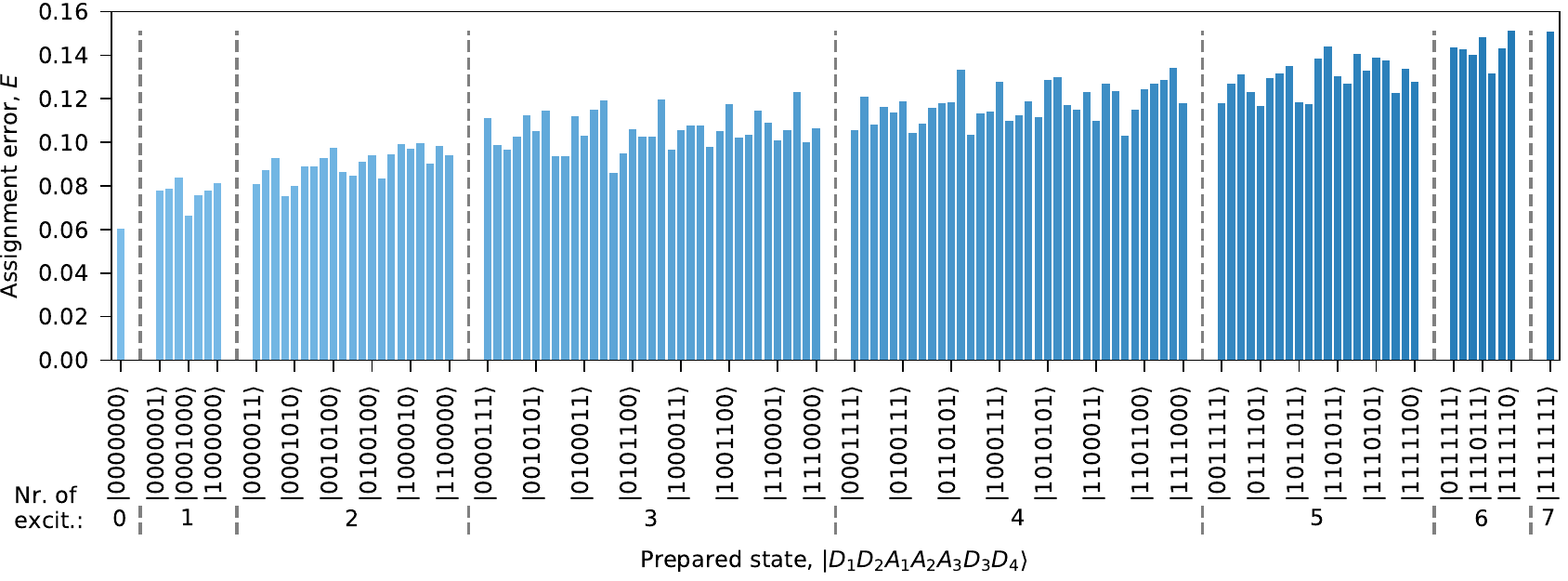}
\caption{Single-shot readout assignment error, $E$, for the simultaneous readout of all 7 qubit.} \label{fig:readout}
\end{figure*}

\begin{figure}
\includegraphics[width=\linewidth]{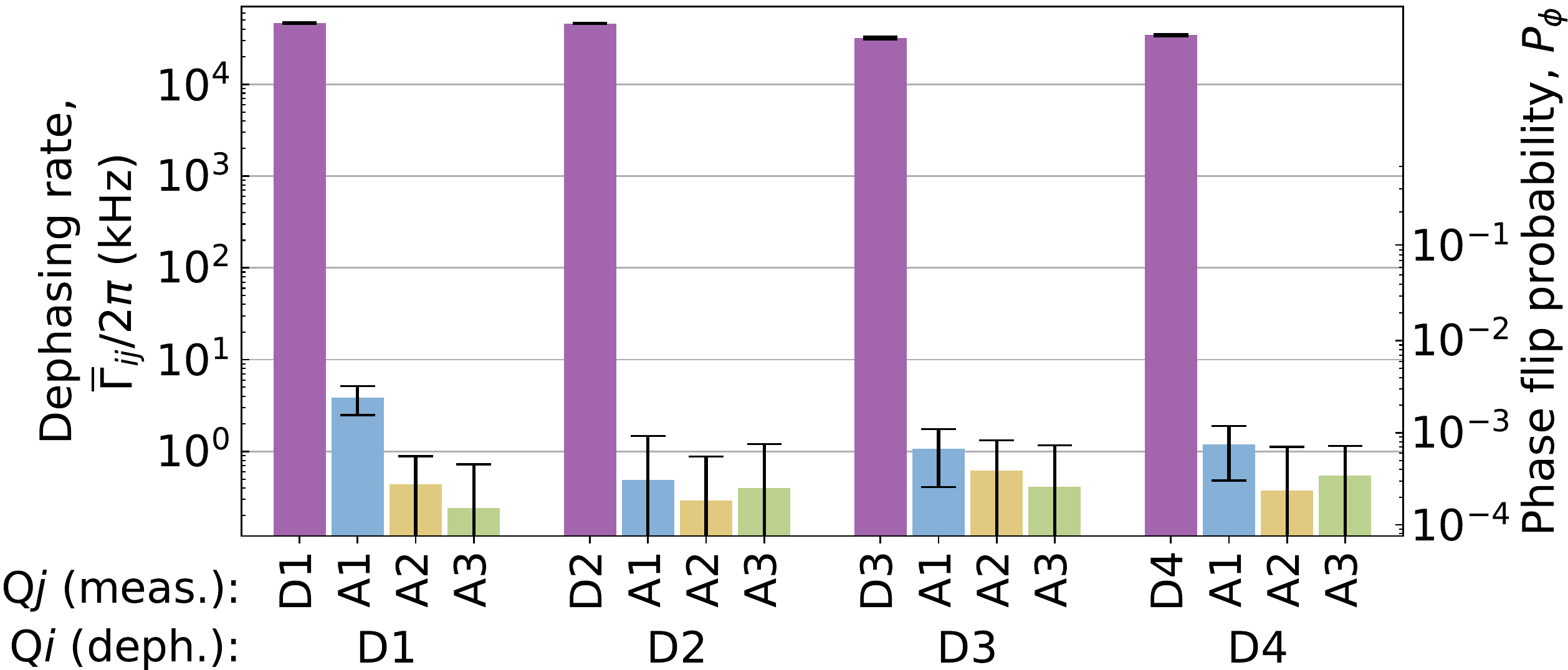}
\caption{Average dephasing rate (left axis) of qubit Q$i$ during a 200~ns long readout pulse on qubit Q$j$ and the corresponding probability of a phase error (right axis) on qubit Q$i$.  } \label{fig:dephase}
\end{figure}

We perform multiplexed readout as detailed in Ref~\cite{Heinsoo2018}. Our readout scheme allows us to selectively address any subset of qubits. The readout is performed with a 200~ns readout pulse and a 300~ns integration window for qubits D1, D2, A1, A2 and A3 and a 300~ns readout pulse with a 400~ns integration window for qubits D3 and D4 due to the smaller dispersive shifts for these qubits. In Fig.~\ref{fig:readout}, we show single-shot readout errors for all computational basis states of the seven qubits with an average assignment error of 11\%.

To characterize measurement induced dephasing on the data qubits when reading out the ancilla qubits, we perform a Ramsey experiment on each of the data qubits. We interleave the Ramsey pulses with a readout on qubit Q$j$~\cite{Heinsoo2018}, and in Fig.~\ref{fig:dephase} we show the resulting additional dephasing rates, $\Gamma_{ij}$, on the data qubits introduced by readout pulses. We can convert the dephasing rates to a probability for introducing a phase error by $P_\phi = [1-\text{exp}(-\Gamma_{ij}\tau_r)]/2$, where $\tau_r$ is the readout time. We find that measurements of the ancilla qubits induce less than 0.3\% phase error on any data qubits.

\begin{figure*}
\includegraphics[width=\linewidth]{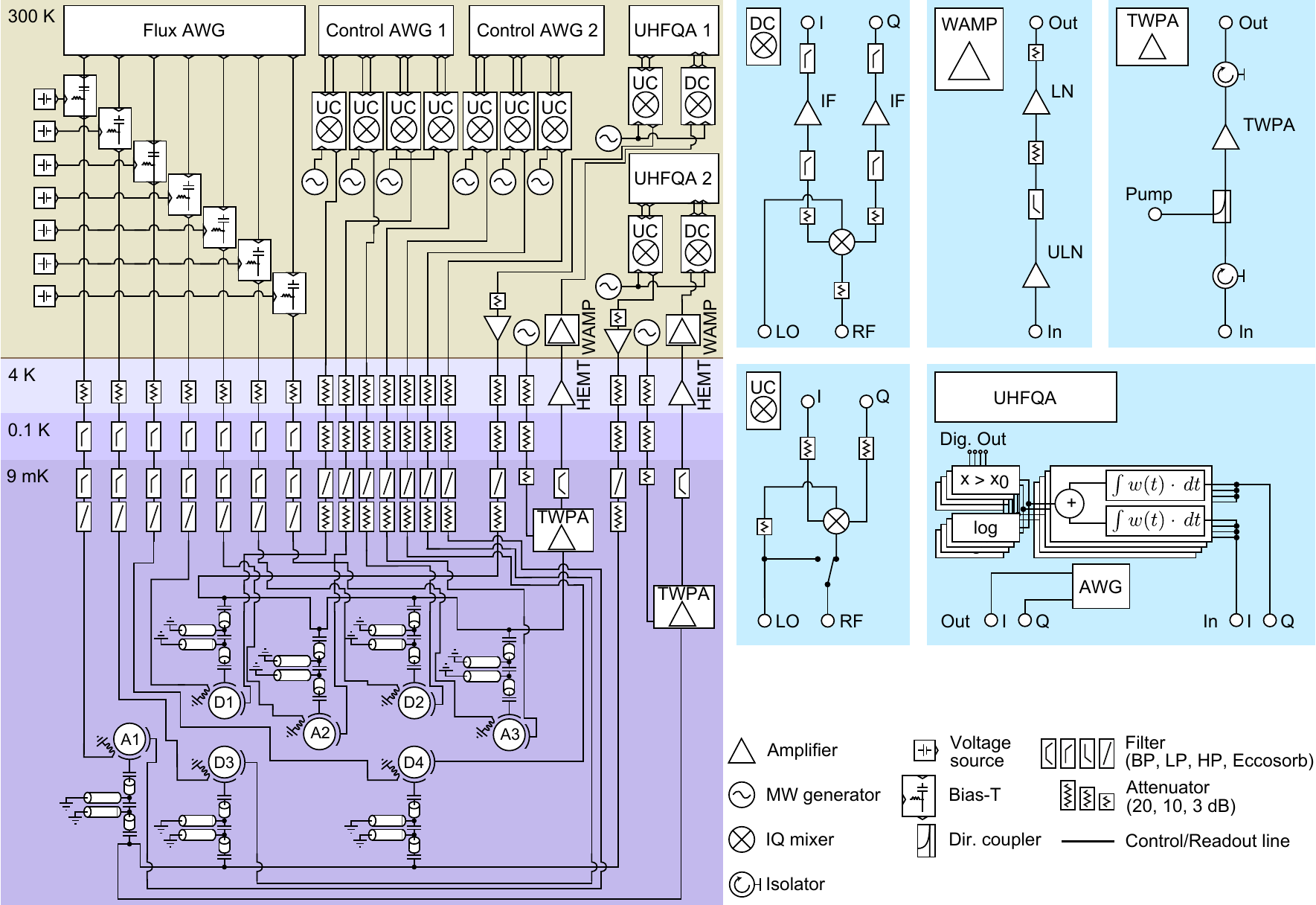}
\caption{Experimental setup described in the text.} \label{fig:setup}
\end{figure*}

\section{Experimental setup}
\label{app:setup}

The seven qubit device is installed at the base plate of a cryogenics setup~\cite{Krinner2019}, see Fig.~\ref{app:setup}. Here, the qubits (indicated by their labels) are controlled by flux and control AWGs through a series of microwave cables each with attenuators and filters, such as bandpass filters (BP), lowpass filters (LP), high pass fiters (HP) and eccosorb filters, installed as indicated. The flux pulses and microwave drive pulses are generated using arbitrary waveform generators (AWG) with 8 channels and a sampling rate of 2.4~GSa/s. The flux pulses are combined with a DC current using a bias-tee. The baseband microwave control pulses are generated at an intermediate frequency (IF) of $100$~MHz and then upconverted to microwave frequencies using IQ mixers installed on upconversion boards (UC). The multiplexed readout pulses, see also Appendix~\ref{app:readout}, are generated and detected using an FPGA based control system (\emph{Zurich Instruments UHFQA}) with a sampling rate of 1.8~GSa/s. The measurement signals at the outputs of the sample are amplified using a wide bandwidth near-quantum-limited traveling wave parametric amplifier (TWPA)~\cite{Macklin2015} connected to isolators at its input and output. Moreover, we installed bandpass filters in the output lines to suppress amplifier noise outside the bandwidth of interest. The output signals are further amplified by high-electron-mobility transistor (HEMT) amplifiers and additional amplifiers at room temperature (WAMP). After amplification, the signals are downconverted (DC) and processed using the weighted integration units of the UHFQAs.

\section{Numerical Simulations}
\label{app:sim}

We model the dynamics of our seven qubit quantum system by a master equation given by
\begin{align}
\dot{\rho} = -\frac{i}{\hbar} [ H(t), \rho ]  + \sum_i \Big[ \hat{c}_i \rho \hat{c}_i\dag - \frac{1}{2} \Big( \hat{c}_i\dag\hat{c}_i \rho + \rho \hat{c}_i\dag \hat{c}_i \Big) \Big], \label{eq:mastereq}
\end{align}
where $\rho$ is the density matrix describing the system at time $t$ and $H(t)$ is the Hamiltonian, the time-dependence of which models the applied gate sequence. The collapse operators $\hat{c}_i$ model incoherent processes. We solve the master equation numerically~\cite{Johansson2013a}. To simplify the description of the system's time evolution, we consider the Hamiltonian to be piece-wise constant, see details in Ref.~\cite{Andersen2019}. In addition, we include the Hamiltonian
\begin{align*}
H_{ZZ}/\hbar &= \sum_{i,j} \alpha_{i,j} \ket{11}_{i,j}\bra{11}
\end{align*}
modeling the residual $ZZ$ coupling $\alpha_{ZZ}$ shown in Fig.~\ref{fig:resid_zz}.
The incoherent errors are described by the Lindblad terms in Eq.~\eqref{eq:mastereq} with
\begin{align*}
\hat{c}_{T_1,i} &= \sqrt{\frac{1}{T_{1,i}}} \sigma_{-,i}, \\
\hat{c}_{T_{\phi,i}} &=  \sqrt{ \frac{1}{2} \Big( \frac{1}{T_{2,i}} - \frac{1}{2T_{1,i}} \Big) } \sigma_{z,i},
\end{align*}
where $T_{1,i}$ and $T_{2,i}$ are the lifetime and decoherence time (Ramsey decay time) of qubit $i$.

To simulate the ancilla measurement, we consider the POVM operators:
\begin{align}
M_0 &= \sqrt{P(0|0)} \ket{0}\bra{0}_A + \sqrt{P(0|1)} \ket{1}\bra{1}_A, \\
M_{1} &= \sqrt{P(1|0)} \ket{0}\bra{0}_A + \sqrt{P(1|1)} \ket{1}\bra{1}_A,
\end{align}
for the outcomes 0 and 1 respectively, where $P(i|j)$ are the experimentally determined probabilities for measuring the state $i$ when preparing the state $j$. We choose for simplicity the POVM operators corresponding to minimal disturbance measurements~\cite{Wiseman2010} as these POVM operators will mostly remove coherences similar to the real physical measurements. We evaluate the probability for each ancilla measurement outcome by $p_i = \text{Tr}(M_i \rho M_i\dag)$ for $i\in\lbrace0,1\rbrace$. The resulting density matrix given a certain measurement outcome $i$ is calculated as $\rho \rightarrow M_i \rho M_i\dag / p_i$.

\end{appendix}

\bibliography{QudevRefDB}

\begin{thebibliography}{52}%
\makeatletter
\providecommand \@ifxundefined [1]{%
 \@ifx{#1\undefined}
}%
\providecommand \@ifnum [1]{%
 \ifnum #1\expandafter \@firstoftwo
 \else \expandafter \@secondoftwo
 \fi
}%
\providecommand \@ifx [1]{%
 \ifx #1\expandafter \@firstoftwo
 \else \expandafter \@secondoftwo
 \fi
}%
\providecommand \natexlab [1]{#1}%
\providecommand \enquote  [1]{``#1''}%
\providecommand \bibnamefont  [1]{#1}%
\providecommand \bibfnamefont [1]{#1}%
\providecommand \citenamefont [1]{#1}%
\providecommand \href@noop [0]{\@secondoftwo}%
\providecommand \href [0]{\begingroup \@sanitize@url \@href}%
\providecommand \@href[1]{\@@startlink{#1}\@@href}%
\providecommand \@@href[1]{\endgroup#1\@@endlink}%
\providecommand \@sanitize@url [0]{\catcode `\\12\catcode `\$12\catcode
  `\&12\catcode `\#12\catcode `\^12\catcode `\_12\catcode `\%12\relax}%
\providecommand \@@startlink[1]{}%
\providecommand \@@endlink[0]{}%
\providecommand \url  [0]{\begingroup\@sanitize@url \@url }%
\providecommand \@url [1]{\endgroup\@href {#1}{\urlprefix }}%
\providecommand \urlprefix  [0]{URL }%
\providecommand \Eprint [0]{\href }%
\providecommand \doibase [0]{http://dx.doi.org/}%
\providecommand \selectlanguage [0]{\@gobble}%
\providecommand \bibinfo  [0]{\@secondoftwo}%
\providecommand \bibfield  [0]{\@secondoftwo}%
\providecommand \translation [1]{[#1]}%
\providecommand \BibitemOpen [0]{}%
\providecommand \bibitemStop [0]{}%
\providecommand \bibitemNoStop [0]{.\EOS\space}%
\providecommand \EOS [0]{\spacefactor3000\relax}%
\providecommand \BibitemShut  [1]{\csname bibitem#1\endcsname}%
\let\auto@bib@innerbib\@empty
\bibitem [{\citenamefont {Zhang}\ \emph {et~al.}(2017)\citenamefont {Zhang},
  \citenamefont {Pagano}, \citenamefont {Hess}, \citenamefont {Kyprianidis},
  \citenamefont {Becker}, \citenamefont {Kaplan}, \citenamefont {Gorshkov},
  \citenamefont {Gong},\ and\ \citenamefont {Monroe}}]{Zhang2017k}%
  \BibitemOpen
  \bibfield  {author} {\bibinfo {author} {\bibfnamefont {J.}~\bibnamefont
  {Zhang}}, \bibinfo {author} {\bibfnamefont {G.}~\bibnamefont {Pagano}},
  \bibinfo {author} {\bibfnamefont {P.~W.}\ \bibnamefont {Hess}}, \bibinfo
  {author} {\bibfnamefont {A.}~\bibnamefont {Kyprianidis}}, \bibinfo {author}
  {\bibfnamefont {P.}~\bibnamefont {Becker}}, \bibinfo {author} {\bibfnamefont
  {H.}~\bibnamefont {Kaplan}}, \bibinfo {author} {\bibfnamefont {A.~V.}\
  \bibnamefont {Gorshkov}}, \bibinfo {author} {\bibfnamefont {Z.-X.}\
  \bibnamefont {Gong}}, \ and\ \bibinfo {author} {\bibfnamefont
  {C.}~\bibnamefont {Monroe}},\ }\bibfield  {title} {\enquote {\bibinfo {title}
  {Observation of a many-body dynamical phase transition with a 53-qubit
  quantum simulator},}\ }\href {http://dx.doi.org/10.1038/nature24654}
  {\bibfield  {journal} {\bibinfo  {journal} {Nature}\ }\textbf {\bibinfo
  {volume} {551}},\ \bibinfo {pages} {601} (\bibinfo {year}
  {2017})}\BibitemShut {NoStop}%
\bibitem [{\citenamefont {Bernien}\ \emph {et~al.}(2017)\citenamefont
  {Bernien}, \citenamefont {Schwartz}, \citenamefont {Keesling}, \citenamefont
  {Levine}, \citenamefont {Omran}, \citenamefont {Pichler}, \citenamefont
  {Choi}, \citenamefont {Zibrov}, \citenamefont {Endres}, \citenamefont
  {Greiner}, \citenamefont {Vuleti\'c},\ and\ \citenamefont
  {Lukin}}]{Bernien2017}%
  \BibitemOpen
  \bibfield  {author} {\bibinfo {author} {\bibfnamefont {Hannes}\ \bibnamefont
  {Bernien}}, \bibinfo {author} {\bibfnamefont {Sylvain}\ \bibnamefont
  {Schwartz}}, \bibinfo {author} {\bibfnamefont {Alexander}\ \bibnamefont
  {Keesling}}, \bibinfo {author} {\bibfnamefont {Harry}\ \bibnamefont
  {Levine}}, \bibinfo {author} {\bibfnamefont {Ahmed}\ \bibnamefont {Omran}},
  \bibinfo {author} {\bibfnamefont {Hannes}\ \bibnamefont {Pichler}}, \bibinfo
  {author} {\bibfnamefont {Soonwon}\ \bibnamefont {Choi}}, \bibinfo {author}
  {\bibfnamefont {Alexander~S.}\ \bibnamefont {Zibrov}}, \bibinfo {author}
  {\bibfnamefont {Manuel}\ \bibnamefont {Endres}}, \bibinfo {author}
  {\bibfnamefont {Markus}\ \bibnamefont {Greiner}}, \bibinfo {author}
  {\bibfnamefont {Vladan}\ \bibnamefont {Vuleti\'c}}, \ and\ \bibinfo {author}
  {\bibfnamefont {Mikhail~D.}\ \bibnamefont {Lukin}},\ }\bibfield  {title}
  {\enquote {\bibinfo {title} {Probing many-body dynamics on a 51-atom quantum
  simulator},}\ }\href {http://dx.doi.org/10.1038/nature24622} {\bibfield
  {journal} {\bibinfo  {journal} {Nature}\ }\textbf {\bibinfo {volume} {551}},\
  \bibinfo {pages} {579--584} (\bibinfo {year} {2017})}\BibitemShut {NoStop}%
\bibitem [{\citenamefont {Arute}\ \emph {et~al.}(2019)\citenamefont {Arute},
  \citenamefont {Arya}, \citenamefont {Babbush}, \citenamefont {Bacon},
  \citenamefont {Bardin}, \citenamefont {Barends}, \citenamefont {Biswas},
  \citenamefont {Boixo}, \citenamefont {Brandao}, \citenamefont {Buell},
  \citenamefont {Burkett}, \citenamefont {Chen}, \citenamefont {Chen},
  \citenamefont {Chiaro}, \citenamefont {Collins}, \citenamefont {Courtney},
  \citenamefont {Dunsworth}, \citenamefont {Farhi}, \citenamefont {Foxen},
  \citenamefont {Fowler}, \citenamefont {Gidney}, \citenamefont {Giustina},
  \citenamefont {Graff}, \citenamefont {Guerin}, \citenamefont {Habegger},
  \citenamefont {Harrigan}, \citenamefont {Hartmann}, \citenamefont {Ho},
  \citenamefont {Hoffmann}, \citenamefont {Huang}, \citenamefont {Humble},
  \citenamefont {Isakov}, \citenamefont {Jeffrey}, \citenamefont {Jiang},
  \citenamefont {Kafri}, \citenamefont {Kechedzhi}, \citenamefont {Kelly},
  \citenamefont {Klimov}, \citenamefont {Knysh}, \citenamefont {Korotkov},
  \citenamefont {Kostritsa}, \citenamefont {Landhuis}, \citenamefont
  {Lindmark}, \citenamefont {Lucero}, \citenamefont {Lyakh}, \citenamefont
  {Mandrà}, \citenamefont {McClean}, \citenamefont {McEwen}, \citenamefont
  {Megrant}, \citenamefont {Mi}, \citenamefont {Michielsen}, \citenamefont
  {Mohseni}, \citenamefont {Mutus}, \citenamefont {Naaman}, \citenamefont
  {Neeley}, \citenamefont {Neill}, \citenamefont {Niu}, \citenamefont {Ostby},
  \citenamefont {Petukhov}, \citenamefont {Platt}, \citenamefont {Quintana},
  \citenamefont {Rieffel}, \citenamefont {Roushan}, \citenamefont {Rubin},
  \citenamefont {Sank}, \citenamefont {Satzinger}, \citenamefont {Smelyanskiy},
  \citenamefont {Sung}, \citenamefont {Trevithick}, \citenamefont
  {Vainsencher}, \citenamefont {Villalonga}, \citenamefont {White},
  \citenamefont {Yao}, \citenamefont {Yeh}, \citenamefont {Zalcman},
  \citenamefont {Neven},\ and\ \citenamefont {Martinis}}]{Arute2019}%
  \BibitemOpen
  \bibfield  {author} {\bibinfo {author} {\bibfnamefont {Frank}\ \bibnamefont
  {Arute}}, \bibinfo {author} {\bibfnamefont {Kunal}\ \bibnamefont {Arya}},
  \bibinfo {author} {\bibfnamefont {Ryan}\ \bibnamefont {Babbush}}, \bibinfo
  {author} {\bibfnamefont {Dave}\ \bibnamefont {Bacon}}, \bibinfo {author}
  {\bibfnamefont {Joseph~C.}\ \bibnamefont {Bardin}}, \bibinfo {author}
  {\bibfnamefont {Rami}\ \bibnamefont {Barends}}, \bibinfo {author}
  {\bibfnamefont {Rupak}\ \bibnamefont {Biswas}}, \bibinfo {author}
  {\bibfnamefont {Sergio}\ \bibnamefont {Boixo}}, \bibinfo {author}
  {\bibfnamefont {Fernando G. S.~L.}\ \bibnamefont {Brandao}}, \bibinfo
  {author} {\bibfnamefont {David~A.}\ \bibnamefont {Buell}}, \bibinfo {author}
  {\bibfnamefont {Brian}\ \bibnamefont {Burkett}}, \bibinfo {author}
  {\bibfnamefont {Yu}~\bibnamefont {Chen}}, \bibinfo {author} {\bibfnamefont
  {Zijun}\ \bibnamefont {Chen}}, \bibinfo {author} {\bibfnamefont {Ben}\
  \bibnamefont {Chiaro}}, \bibinfo {author} {\bibfnamefont {Roberto}\
  \bibnamefont {Collins}}, \bibinfo {author} {\bibfnamefont {William}\
  \bibnamefont {Courtney}}, \bibinfo {author} {\bibfnamefont {Andrew}\
  \bibnamefont {Dunsworth}}, \bibinfo {author} {\bibfnamefont {Edward}\
  \bibnamefont {Farhi}}, \bibinfo {author} {\bibfnamefont {Brooks}\
  \bibnamefont {Foxen}}, \bibinfo {author} {\bibfnamefont {Austin}\
  \bibnamefont {Fowler}}, \bibinfo {author} {\bibfnamefont {Craig}\
  \bibnamefont {Gidney}}, \bibinfo {author} {\bibfnamefont {Marissa}\
  \bibnamefont {Giustina}}, \bibinfo {author} {\bibfnamefont {Rob}\
  \bibnamefont {Graff}}, \bibinfo {author} {\bibfnamefont {Keith}\ \bibnamefont
  {Guerin}}, \bibinfo {author} {\bibfnamefont {Steve}\ \bibnamefont
  {Habegger}}, \bibinfo {author} {\bibfnamefont {Matthew~P.}\ \bibnamefont
  {Harrigan}}, \bibinfo {author} {\bibfnamefont {Michael~J.}\ \bibnamefont
  {Hartmann}}, \bibinfo {author} {\bibfnamefont {Alan}\ \bibnamefont {Ho}},
  \bibinfo {author} {\bibfnamefont {Markus}\ \bibnamefont {Hoffmann}}, \bibinfo
  {author} {\bibfnamefont {Trent}\ \bibnamefont {Huang}}, \bibinfo {author}
  {\bibfnamefont {Travis~S.}\ \bibnamefont {Humble}}, \bibinfo {author}
  {\bibfnamefont {Sergei~V.}\ \bibnamefont {Isakov}}, \bibinfo {author}
  {\bibfnamefont {Evan}\ \bibnamefont {Jeffrey}}, \bibinfo {author}
  {\bibfnamefont {Zhang}\ \bibnamefont {Jiang}}, \bibinfo {author}
  {\bibfnamefont {Dvir}\ \bibnamefont {Kafri}}, \bibinfo {author}
  {\bibfnamefont {Kostyantyn}\ \bibnamefont {Kechedzhi}}, \bibinfo {author}
  {\bibfnamefont {Julian}\ \bibnamefont {Kelly}}, \bibinfo {author}
  {\bibfnamefont {Paul~V.}\ \bibnamefont {Klimov}}, \bibinfo {author}
  {\bibfnamefont {Sergey}\ \bibnamefont {Knysh}}, \bibinfo {author}
  {\bibfnamefont {Alexander}\ \bibnamefont {Korotkov}}, \bibinfo {author}
  {\bibfnamefont {Fedor}\ \bibnamefont {Kostritsa}}, \bibinfo {author}
  {\bibfnamefont {David}\ \bibnamefont {Landhuis}}, \bibinfo {author}
  {\bibfnamefont {Mike}\ \bibnamefont {Lindmark}}, \bibinfo {author}
  {\bibfnamefont {Erik}\ \bibnamefont {Lucero}}, \bibinfo {author}
  {\bibfnamefont {Dmitry}\ \bibnamefont {Lyakh}}, \bibinfo {author}
  {\bibfnamefont {Salvatore}\ \bibnamefont {Mandrà}}, \bibinfo {author}
  {\bibfnamefont {Jarrod~R.}\ \bibnamefont {McClean}}, \bibinfo {author}
  {\bibfnamefont {Matthew}\ \bibnamefont {McEwen}}, \bibinfo {author}
  {\bibfnamefont {Anthony}\ \bibnamefont {Megrant}}, \bibinfo {author}
  {\bibfnamefont {Xiao}\ \bibnamefont {Mi}}, \bibinfo {author} {\bibfnamefont
  {Kristel}\ \bibnamefont {Michielsen}}, \bibinfo {author} {\bibfnamefont
  {Masoud}\ \bibnamefont {Mohseni}}, \bibinfo {author} {\bibfnamefont {Josh}\
  \bibnamefont {Mutus}}, \bibinfo {author} {\bibfnamefont {Ofer}\ \bibnamefont
  {Naaman}}, \bibinfo {author} {\bibfnamefont {Matthew}\ \bibnamefont
  {Neeley}}, \bibinfo {author} {\bibfnamefont {Charles}\ \bibnamefont {Neill}},
  \bibinfo {author} {\bibfnamefont {Murphy~Yuezhen}\ \bibnamefont {Niu}},
  \bibinfo {author} {\bibfnamefont {Eric}\ \bibnamefont {Ostby}}, \bibinfo
  {author} {\bibfnamefont {Andre}\ \bibnamefont {Petukhov}}, \bibinfo {author}
  {\bibfnamefont {John~C.}\ \bibnamefont {Platt}}, \bibinfo {author}
  {\bibfnamefont {Chris}\ \bibnamefont {Quintana}}, \bibinfo {author}
  {\bibfnamefont {Eleanor~G.}\ \bibnamefont {Rieffel}}, \bibinfo {author}
  {\bibfnamefont {Pedram}\ \bibnamefont {Roushan}}, \bibinfo {author}
  {\bibfnamefont {Nicholas~C.}\ \bibnamefont {Rubin}}, \bibinfo {author}
  {\bibfnamefont {Daniel}\ \bibnamefont {Sank}}, \bibinfo {author}
  {\bibfnamefont {Kevin~J.}\ \bibnamefont {Satzinger}}, \bibinfo {author}
  {\bibfnamefont {Vadim}\ \bibnamefont {Smelyanskiy}}, \bibinfo {author}
  {\bibfnamefont {Kevin~J.}\ \bibnamefont {Sung}}, \bibinfo {author}
  {\bibfnamefont {Matthew~D.}\ \bibnamefont {Trevithick}}, \bibinfo {author}
  {\bibfnamefont {Amit}\ \bibnamefont {Vainsencher}}, \bibinfo {author}
  {\bibfnamefont {Benjamin}\ \bibnamefont {Villalonga}}, \bibinfo {author}
  {\bibfnamefont {Theodore}\ \bibnamefont {White}}, \bibinfo {author}
  {\bibfnamefont {Z.~Jamie}\ \bibnamefont {Yao}}, \bibinfo {author}
  {\bibfnamefont {Ping}\ \bibnamefont {Yeh}}, \bibinfo {author} {\bibfnamefont
  {Adam}\ \bibnamefont {Zalcman}}, \bibinfo {author} {\bibfnamefont {Hartmut}\
  \bibnamefont {Neven}}, \ and\ \bibinfo {author} {\bibfnamefont {John~M.}\
  \bibnamefont {Martinis}},\ }\bibfield  {title} {\enquote {\bibinfo {title}
  {Quantum supremacy using a programmable superconducting processor},}\ }\href
  {\doibase doi:10.1038/s41586-019-1666-5} {\bibfield  {journal} {\bibinfo
  {journal} {Nature}\ }\textbf {\bibinfo {volume} {574}},\ \bibinfo {pages}
  {505--510} (\bibinfo {year} {2019})}\BibitemShut {NoStop}%
\bibitem [{\citenamefont {Preskill}(2018)}]{Preskill2018}%
  \BibitemOpen
  \bibfield  {author} {\bibinfo {author} {\bibfnamefont {John}\ \bibnamefont
  {Preskill}},\ }\bibfield  {title} {\enquote {\bibinfo {title} {Quantum
  {C}omputing in the {NISQ} era and beyond},}\ }\href {\doibase
  10.22331/q-2018-08-06-79} {\bibfield  {journal} {\bibinfo  {journal}
  {{Quantum}}\ }\textbf {\bibinfo {volume} {2}},\ \bibinfo {pages} {79}
  (\bibinfo {year} {2018})}\BibitemShut {NoStop}%
\bibitem [{\citenamefont {Cory}\ \emph {et~al.}(1998)\citenamefont {Cory},
  \citenamefont {Price}, \citenamefont {Maas}, \citenamefont {Knill},
  \citenamefont {Laflamme}, \citenamefont {Zurek}, \citenamefont {Havel},\ and\
  \citenamefont {Somaroo}}]{Cory1998}%
  \BibitemOpen
  \bibfield  {author} {\bibinfo {author} {\bibfnamefont {D.~G.}\ \bibnamefont
  {Cory}}, \bibinfo {author} {\bibfnamefont {M.~D.}\ \bibnamefont {Price}},
  \bibinfo {author} {\bibfnamefont {W.}~\bibnamefont {Maas}}, \bibinfo {author}
  {\bibfnamefont {E.}~\bibnamefont {Knill}}, \bibinfo {author} {\bibfnamefont
  {R.}~\bibnamefont {Laflamme}}, \bibinfo {author} {\bibfnamefont {W.~H.}\
  \bibnamefont {Zurek}}, \bibinfo {author} {\bibfnamefont {T.~F.}\ \bibnamefont
  {Havel}}, \ and\ \bibinfo {author} {\bibfnamefont {S.~S.}\ \bibnamefont
  {Somaroo}},\ }\bibfield  {title} {\enquote {\bibinfo {title} {Experimental
  quantum error correction},}\ }\href {\doibase 10.1103/PhysRevLett.81.2152}
  {\bibfield  {journal} {\bibinfo  {journal} {Phys. Rev. Lett.}\ }\textbf
  {\bibinfo {volume} {81}},\ \bibinfo {pages} {2152--2155} (\bibinfo {year}
  {1998})}\BibitemShut {NoStop}%
\bibitem [{\citenamefont {Chiaverini}\ \emph {et~al.}(2004)\citenamefont
  {Chiaverini}, \citenamefont {Leibfried}, \citenamefont {Schaetz},
  \citenamefont {Barrett}, \citenamefont {Blakestad}, \citenamefont {Britton},
  \citenamefont {Itano}, \citenamefont {Jost}, \citenamefont {Knill},
  \citenamefont {Langer}, \citenamefont {Ozeri},\ and\ \citenamefont
  {Wineland}}]{Chiaverini2004}%
  \BibitemOpen
  \bibfield  {author} {\bibinfo {author} {\bibfnamefont {J.}~\bibnamefont
  {Chiaverini}}, \bibinfo {author} {\bibfnamefont {D.}~\bibnamefont
  {Leibfried}}, \bibinfo {author} {\bibfnamefont {T.}~\bibnamefont {Schaetz}},
  \bibinfo {author} {\bibfnamefont {M.~D.}\ \bibnamefont {Barrett}}, \bibinfo
  {author} {\bibfnamefont {R.~B.}\ \bibnamefont {Blakestad}}, \bibinfo {author}
  {\bibfnamefont {J.}~\bibnamefont {Britton}}, \bibinfo {author} {\bibfnamefont
  {W.~M.}\ \bibnamefont {Itano}}, \bibinfo {author} {\bibfnamefont {J.~D.}\
  \bibnamefont {Jost}}, \bibinfo {author} {\bibfnamefont {E.}~\bibnamefont
  {Knill}}, \bibinfo {author} {\bibfnamefont {C.}~\bibnamefont {Langer}},
  \bibinfo {author} {\bibfnamefont {R.}~\bibnamefont {Ozeri}}, \ and\ \bibinfo
  {author} {\bibfnamefont {D.~J.}\ \bibnamefont {Wineland}},\ }\bibfield
  {title} {\enquote {\bibinfo {title} {Realization of quantum error
  correction},}\ }\href {\doibase 10.1038/nature03074} {\bibfield  {journal}
  {\bibinfo  {journal} {Nature}\ }\textbf {\bibinfo {volume} {432}},\ \bibinfo
  {pages} {602--605} (\bibinfo {year} {2004})}\BibitemShut {NoStop}%
\bibitem [{\citenamefont {Schindler}\ \emph {et~al.}(2011)\citenamefont
  {Schindler}, \citenamefont {Barreiro}, \citenamefont {Monz}, \citenamefont
  {Nebendahl}, \citenamefont {Nigg}, \citenamefont {Chwalla}, \citenamefont
  {Hennrich},\ and\ \citenamefont {Blatt}}]{Schindler2011}%
  \BibitemOpen
  \bibfield  {author} {\bibinfo {author} {\bibfnamefont {Philipp}\ \bibnamefont
  {Schindler}}, \bibinfo {author} {\bibfnamefont {Julio~T.}\ \bibnamefont
  {Barreiro}}, \bibinfo {author} {\bibfnamefont {Thomas}\ \bibnamefont {Monz}},
  \bibinfo {author} {\bibfnamefont {Volckmar}\ \bibnamefont {Nebendahl}},
  \bibinfo {author} {\bibfnamefont {Daniel}\ \bibnamefont {Nigg}}, \bibinfo
  {author} {\bibfnamefont {Michael}\ \bibnamefont {Chwalla}}, \bibinfo {author}
  {\bibfnamefont {Markus}\ \bibnamefont {Hennrich}}, \ and\ \bibinfo {author}
  {\bibfnamefont {Rainer}\ \bibnamefont {Blatt}},\ }\bibfield  {title}
  {\enquote {\bibinfo {title} {Experimental repetitive quantum error
  correction},}\ }\href {\doibase 10.1126/science.1203329} {\bibfield
  {journal} {\bibinfo  {journal} {Science}\ }\textbf {\bibinfo {volume}
  {332}},\ \bibinfo {pages} {1059--1061} (\bibinfo {year} {2011})}\BibitemShut
  {NoStop}%
\bibitem [{\citenamefont {Lanyon}\ \emph {et~al.}(2013)\citenamefont {Lanyon},
  \citenamefont {Jurcevic}, \citenamefont {Zwerger}, \citenamefont {Hempel},
  \citenamefont {Martinez}, \citenamefont {D\"ur}, \citenamefont {Briegel},
  \citenamefont {Blatt},\ and\ \citenamefont {Roos}}]{Lanyon2013}%
  \BibitemOpen
  \bibfield  {author} {\bibinfo {author} {\bibfnamefont {B.~P.}\ \bibnamefont
  {Lanyon}}, \bibinfo {author} {\bibfnamefont {P.}~\bibnamefont {Jurcevic}},
  \bibinfo {author} {\bibfnamefont {M.}~\bibnamefont {Zwerger}}, \bibinfo
  {author} {\bibfnamefont {C.}~\bibnamefont {Hempel}}, \bibinfo {author}
  {\bibfnamefont {E.~A.}\ \bibnamefont {Martinez}}, \bibinfo {author}
  {\bibfnamefont {W.}~\bibnamefont {D\"ur}}, \bibinfo {author} {\bibfnamefont
  {H.~J.}\ \bibnamefont {Briegel}}, \bibinfo {author} {\bibfnamefont
  {R.}~\bibnamefont {Blatt}}, \ and\ \bibinfo {author} {\bibfnamefont {C.~F.}\
  \bibnamefont {Roos}},\ }\bibfield  {title} {\enquote {\bibinfo {title}
  {Measurement-based quantum computation with trapped ions},}\ }\href {\doibase
  10.1103/PhysRevLett.111.210501} {\bibfield  {journal} {\bibinfo  {journal}
  {Phys. Rev. Lett.}\ }\textbf {\bibinfo {volume} {111}},\ \bibinfo {pages}
  {210501} (\bibinfo {year} {2013})}\BibitemShut {NoStop}%
\bibitem [{\citenamefont {Linke}\ \emph {et~al.}(2017)\citenamefont {Linke},
  \citenamefont {Gutierrez}, \citenamefont {Landsman}, \citenamefont {Figgatt},
  \citenamefont {Debnath}, \citenamefont {Brown},\ and\ \citenamefont
  {Monroe}}]{Linke2017a}%
  \BibitemOpen
  \bibfield  {author} {\bibinfo {author} {\bibfnamefont {N.~M.}\ \bibnamefont
  {Linke}}, \bibinfo {author} {\bibfnamefont {M.}~\bibnamefont {Gutierrez}},
  \bibinfo {author} {\bibfnamefont {K.~A.}\ \bibnamefont {Landsman}}, \bibinfo
  {author} {\bibfnamefont {C.}~\bibnamefont {Figgatt}}, \bibinfo {author}
  {\bibfnamefont {S.}~\bibnamefont {Debnath}}, \bibinfo {author} {\bibfnamefont
  {K.~R.}\ \bibnamefont {Brown}}, \ and\ \bibinfo {author} {\bibfnamefont
  {C.}~\bibnamefont {Monroe}},\ }\bibfield  {title} {\enquote {\bibinfo {title}
  {Fault-tolerant quantum error detection},}\ }\href {\doibase
  10.1126/sciadv.1701074} {\bibfield  {journal} {\bibinfo  {journal} {Science
  Advances}\ }\textbf {\bibinfo {volume} {3}},\ \bibinfo {pages} {10} (\bibinfo
  {year} {2017})}\BibitemShut {NoStop}%
\bibitem [{\citenamefont {Yao}\ \emph {et~al.}(2012)\citenamefont {Yao},
  \citenamefont {Wang}, \citenamefont {Chen}, \citenamefont {Gao},
  \citenamefont {Fowler}, \citenamefont {Raussendorf}, \citenamefont {Chen},
  \citenamefont {Liu}, \citenamefont {Lu}, \citenamefont {Deng}, \citenamefont
  {Chen},\ and\ \citenamefont {Pan}}]{Yao2012a}%
  \BibitemOpen
  \bibfield  {author} {\bibinfo {author} {\bibfnamefont {Xing-Can}\
  \bibnamefont {Yao}}, \bibinfo {author} {\bibfnamefont {Tian-Xiong}\
  \bibnamefont {Wang}}, \bibinfo {author} {\bibfnamefont {Hao-Ze}\ \bibnamefont
  {Chen}}, \bibinfo {author} {\bibfnamefont {Wei-Bo}\ \bibnamefont {Gao}},
  \bibinfo {author} {\bibfnamefont {Austin~G.}\ \bibnamefont {Fowler}},
  \bibinfo {author} {\bibfnamefont {Robert}\ \bibnamefont {Raussendorf}},
  \bibinfo {author} {\bibfnamefont {Zeng-Bing}\ \bibnamefont {Chen}}, \bibinfo
  {author} {\bibfnamefont {Nai-Le}\ \bibnamefont {Liu}}, \bibinfo {author}
  {\bibfnamefont {Chao-Yang}\ \bibnamefont {Lu}}, \bibinfo {author}
  {\bibfnamefont {You-Jin}\ \bibnamefont {Deng}}, \bibinfo {author}
  {\bibfnamefont {Yu-Ao}\ \bibnamefont {Chen}}, \ and\ \bibinfo {author}
  {\bibfnamefont {Jian-Wei}\ \bibnamefont {Pan}},\ }\bibfield  {title}
  {\enquote {\bibinfo {title} {Experimental demonstration of topological error
  correction},}\ }\href {http://dx.doi.org/10.1038/nature10770} {\bibfield
  {journal} {\bibinfo  {journal} {Nature}\ }\textbf {\bibinfo {volume} {482}},\
  \bibinfo {pages} {489--494} (\bibinfo {year} {2012})}\BibitemShut {NoStop}%
\bibitem [{\citenamefont {Bell}\ \emph {et~al.}(2014)\citenamefont {Bell},
  \citenamefont {Herrera-Marti}, \citenamefont {Tame}, \citenamefont {Markham},
  \citenamefont {Wadsworth},\ and\ \citenamefont {Rarity}}]{Bell2014}%
  \BibitemOpen
  \bibfield  {author} {\bibinfo {author} {\bibfnamefont {B.~A.}\ \bibnamefont
  {Bell}}, \bibinfo {author} {\bibfnamefont {D.~A.}\ \bibnamefont
  {Herrera-Marti}}, \bibinfo {author} {\bibfnamefont {M.~S.}\ \bibnamefont
  {Tame}}, \bibinfo {author} {\bibfnamefont {D.}~\bibnamefont {Markham}},
  \bibinfo {author} {\bibfnamefont {W.~J.}\ \bibnamefont {Wadsworth}}, \ and\
  \bibinfo {author} {\bibfnamefont {J.~G.}\ \bibnamefont {Rarity}},\ }\bibfield
   {title} {\enquote {\bibinfo {title} {Experimental demonstration of a graph
  state quantum error-correction code},}\ }\href {\doibase 10.1038/ncomms4658}
  {\bibfield  {journal} {\bibinfo  {journal} {Nat. Commun.}\ }\textbf {\bibinfo
  {volume} {5}},\ \bibinfo {pages} {3658} (\bibinfo {year} {2014})}\BibitemShut
  {NoStop}%
\bibitem [{\citenamefont {Cramer}\ \emph {et~al.}(2016)\citenamefont {Cramer},
  \citenamefont {Kalb}, \citenamefont {Rol}, \citenamefont {Hensen},
  \citenamefont {Blok}, \citenamefont {Markham}, \citenamefont {Twitchen},
  \citenamefont {Hanson},\ and\ \citenamefont {Taminiau}}]{Cramer2016}%
  \BibitemOpen
  \bibfield  {author} {\bibinfo {author} {\bibfnamefont {J.}~\bibnamefont
  {Cramer}}, \bibinfo {author} {\bibfnamefont {N.}~\bibnamefont {Kalb}},
  \bibinfo {author} {\bibfnamefont {M.~A.}\ \bibnamefont {Rol}}, \bibinfo
  {author} {\bibfnamefont {B.}~\bibnamefont {Hensen}}, \bibinfo {author}
  {\bibfnamefont {M.~S.}\ \bibnamefont {Blok}}, \bibinfo {author}
  {\bibfnamefont {M.}~\bibnamefont {Markham}}, \bibinfo {author} {\bibfnamefont
  {D.~J.}\ \bibnamefont {Twitchen}}, \bibinfo {author} {\bibfnamefont
  {R.}~\bibnamefont {Hanson}}, \ and\ \bibinfo {author} {\bibfnamefont {T.~H.}\
  \bibnamefont {Taminiau}},\ }\bibfield  {title} {\enquote {\bibinfo {title}
  {Repeated quantum error correction on a continuously encoded qubit by
  real-time feedback},}\ }\href {\doibase https://doi.org/10.1038/ncomms11526}
  {\bibfield  {journal} {\bibinfo  {journal} {Nat. Commun.}\ }\textbf {\bibinfo
  {volume} {7}},\ \bibinfo {pages} {11526} (\bibinfo {year}
  {2016})}\BibitemShut {NoStop}%
\bibitem [{\citenamefont {Reed}\ \emph {et~al.}(2012)\citenamefont {Reed},
  \citenamefont {DiCarlo}, \citenamefont {Nigg}, \citenamefont {Sun},
  \citenamefont {Frunzio}, \citenamefont {Girvin},\ and\ \citenamefont
  {Schoelkopf}}]{Reed2012}%
  \BibitemOpen
  \bibfield  {author} {\bibinfo {author} {\bibfnamefont {M.~D.}\ \bibnamefont
  {Reed}}, \bibinfo {author} {\bibfnamefont {L.}~\bibnamefont {DiCarlo}},
  \bibinfo {author} {\bibfnamefont {S.~E.}\ \bibnamefont {Nigg}}, \bibinfo
  {author} {\bibfnamefont {L.}~\bibnamefont {Sun}}, \bibinfo {author}
  {\bibfnamefont {L.}~\bibnamefont {Frunzio}}, \bibinfo {author} {\bibfnamefont
  {S.~M.}\ \bibnamefont {Girvin}}, \ and\ \bibinfo {author} {\bibfnamefont
  {R.~J.}\ \bibnamefont {Schoelkopf}},\ }\bibfield  {title} {\enquote {\bibinfo
  {title} {Realization of three-qubit quantum error correction with
  superconducting circuits},}\ }\href {\doibase 10.1038/nature10786} {\bibfield
   {journal} {\bibinfo  {journal} {Nature}\ }\textbf {\bibinfo {volume}
  {482}},\ \bibinfo {pages} {382--385} (\bibinfo {year} {2012})}\BibitemShut
  {NoStop}%
\bibitem [{\citenamefont {Shankar}\ \emph {et~al.}(2013)\citenamefont
  {Shankar}, \citenamefont {Hatridge}, \citenamefont {Leghtas}, \citenamefont
  {Sliwa}, \citenamefont {Narla}, \citenamefont {Vool}, \citenamefont {Girvin},
  \citenamefont {Frunzio}, \citenamefont {Mirrahimi},\ and\ \citenamefont
  {Devoret}}]{Shankar2013a}%
  \BibitemOpen
  \bibfield  {author} {\bibinfo {author} {\bibfnamefont {S.}~\bibnamefont
  {Shankar}}, \bibinfo {author} {\bibfnamefont {M.}~\bibnamefont {Hatridge}},
  \bibinfo {author} {\bibfnamefont {Z.}~\bibnamefont {Leghtas}}, \bibinfo
  {author} {\bibfnamefont {K.~M.}\ \bibnamefont {Sliwa}}, \bibinfo {author}
  {\bibfnamefont {A.}~\bibnamefont {Narla}}, \bibinfo {author} {\bibfnamefont
  {U.}~\bibnamefont {Vool}}, \bibinfo {author} {\bibfnamefont {S.~M.}\
  \bibnamefont {Girvin}}, \bibinfo {author} {\bibfnamefont {L.}~\bibnamefont
  {Frunzio}}, \bibinfo {author} {\bibfnamefont {M.}~\bibnamefont {Mirrahimi}},
  \ and\ \bibinfo {author} {\bibfnamefont {M.~H.}\ \bibnamefont {Devoret}},\
  }\bibfield  {title} {\enquote {\bibinfo {title} {Autonomously stabilized
  entanglement between two superconducting quantum bits},}\ }\href {\doibase
  10.1038/nature12802} {\bibfield  {journal} {\bibinfo  {journal} {Nature}\
  }\textbf {\bibinfo {volume} {504}},\ \bibinfo {pages} {419--422} (\bibinfo
  {year} {2013})}\BibitemShut {NoStop}%
\bibitem [{\citenamefont {Rist\`e}\ \emph {et~al.}(2015)\citenamefont
  {Rist\`e}, \citenamefont {Poletto}, \citenamefont {Huang}, \citenamefont
  {Bruno}, \citenamefont {Vesterinen}, \citenamefont {Saira},\ and\
  \citenamefont {DiCarlo}}]{Riste2015}%
  \BibitemOpen
  \bibfield  {author} {\bibinfo {author} {\bibfnamefont {D.}~\bibnamefont
  {Rist\`e}}, \bibinfo {author} {\bibfnamefont {S.}~\bibnamefont {Poletto}},
  \bibinfo {author} {\bibfnamefont {M.-Z.}\ \bibnamefont {Huang}}, \bibinfo
  {author} {\bibfnamefont {A.}~\bibnamefont {Bruno}}, \bibinfo {author}
  {\bibfnamefont {V.}~\bibnamefont {Vesterinen}}, \bibinfo {author}
  {\bibfnamefont {O.-P.}\ \bibnamefont {Saira}}, \ and\ \bibinfo {author}
  {\bibfnamefont {L.}~\bibnamefont {DiCarlo}},\ }\bibfield  {title} {\enquote
  {\bibinfo {title} {Detecting bit-flip errors in a logical qubit using
  stabilizer measurements},}\ }\href {http://dx.doi.org/10.1038/ncomms7983}
  {\bibfield  {journal} {\bibinfo  {journal} {Nat. Commun.}\ }\textbf {\bibinfo
  {volume} {6}},\ \bibinfo {pages} {6983} (\bibinfo {year} {2015})}\BibitemShut
  {NoStop}%
\bibitem [{\citenamefont {Kelly}\ \emph {et~al.}(2015)\citenamefont {Kelly},
  \citenamefont {Barends}, \citenamefont {Fowler}, \citenamefont {Megrant},
  \citenamefont {Jeffrey}, \citenamefont {White}, \citenamefont {Sank},
  \citenamefont {Mutus}, \citenamefont {Campbell}, \citenamefont {Chen},
  \citenamefont {Chen}, \citenamefont {Chiaro}, \citenamefont {Dunsworth},
  \citenamefont {Hoi}, \citenamefont {Neill}, \citenamefont {O'Malley},
  \citenamefont {Quintana}, \citenamefont {Roushan}, \citenamefont
  {Vainsencher}, \citenamefont {Wenner}, \citenamefont {Cleland},\ and\
  \citenamefont {Martinis}}]{Kelly2015}%
  \BibitemOpen
  \bibfield  {author} {\bibinfo {author} {\bibfnamefont {J.}~\bibnamefont
  {Kelly}}, \bibinfo {author} {\bibfnamefont {R.}~\bibnamefont {Barends}},
  \bibinfo {author} {\bibfnamefont {A.~G.}\ \bibnamefont {Fowler}}, \bibinfo
  {author} {\bibfnamefont {A.}~\bibnamefont {Megrant}}, \bibinfo {author}
  {\bibfnamefont {E.}~\bibnamefont {Jeffrey}}, \bibinfo {author} {\bibfnamefont
  {T.~C.}\ \bibnamefont {White}}, \bibinfo {author} {\bibfnamefont
  {D.}~\bibnamefont {Sank}}, \bibinfo {author} {\bibfnamefont {J.~Y.}\
  \bibnamefont {Mutus}}, \bibinfo {author} {\bibfnamefont {B.}~\bibnamefont
  {Campbell}}, \bibinfo {author} {\bibfnamefont {Y.}~\bibnamefont {Chen}},
  \bibinfo {author} {\bibfnamefont {Z.}~\bibnamefont {Chen}}, \bibinfo {author}
  {\bibfnamefont {B.}~\bibnamefont {Chiaro}}, \bibinfo {author} {\bibfnamefont
  {A.}~\bibnamefont {Dunsworth}}, \bibinfo {author} {\bibfnamefont {I.-C.}\
  \bibnamefont {Hoi}}, \bibinfo {author} {\bibfnamefont {C.}~\bibnamefont
  {Neill}}, \bibinfo {author} {\bibfnamefont {P.~J.~J.}\ \bibnamefont
  {O'Malley}}, \bibinfo {author} {\bibfnamefont {C.}~\bibnamefont {Quintana}},
  \bibinfo {author} {\bibfnamefont {P.}~\bibnamefont {Roushan}}, \bibinfo
  {author} {\bibfnamefont {A.}~\bibnamefont {Vainsencher}}, \bibinfo {author}
  {\bibfnamefont {J.}~\bibnamefont {Wenner}}, \bibinfo {author} {\bibfnamefont
  {A.~N.}\ \bibnamefont {Cleland}}, \ and\ \bibinfo {author} {\bibfnamefont
  {J.~M.}\ \bibnamefont {Martinis}},\ }\bibfield  {title} {\enquote {\bibinfo
  {title} {State preservation by repetitive error detection in a
  superconducting quantum circuit},}\ }\href {\doibase doi:10.1038/nature14270}
  {\bibfield  {journal} {\bibinfo  {journal} {Nature}\ }\textbf {\bibinfo
  {volume} {519}},\ \bibinfo {pages} {66} (\bibinfo {year} {2015})}\BibitemShut
  {NoStop}%
\bibitem [{\citenamefont {C\'{o}rcoles}\ \emph {et~al.}(2015)\citenamefont
  {C\'{o}rcoles}, \citenamefont {Magesan}, \citenamefont {Srinivasan},
  \citenamefont {Cross}, \citenamefont {Steffen}, \citenamefont {Gambetta},\
  and\ \citenamefont {Chow}}]{Corcoles2015}%
  \BibitemOpen
  \bibfield  {author} {\bibinfo {author} {\bibfnamefont {A.~D.}\ \bibnamefont
  {C\'{o}rcoles}}, \bibinfo {author} {\bibfnamefont {Easwar}\ \bibnamefont
  {Magesan}}, \bibinfo {author} {\bibfnamefont {Srikanth~J.}\ \bibnamefont
  {Srinivasan}}, \bibinfo {author} {\bibfnamefont {Andrew~W.}\ \bibnamefont
  {Cross}}, \bibinfo {author} {\bibfnamefont {M.}~\bibnamefont {Steffen}},
  \bibinfo {author} {\bibfnamefont {Jay~M.}\ \bibnamefont {Gambetta}}, \ and\
  \bibinfo {author} {\bibfnamefont {Jerry~M.}\ \bibnamefont {Chow}},\
  }\bibfield  {title} {\enquote {\bibinfo {title} {Demonstration of a quantum
  error detection code using a square lattice of four superconducting
  qubits},}\ }\href {http://dx.doi.org/10.1038/ncomms7979} {\bibfield
  {journal} {\bibinfo  {journal} {Nat. Commun.}\ }\textbf {\bibinfo {volume}
  {6}},\ \bibinfo {pages} {6979} (\bibinfo {year} {2015})}\BibitemShut
  {NoStop}%
\bibitem [{\citenamefont {Ofek}\ \emph {et~al.}(2016)\citenamefont {Ofek},
  \citenamefont {Petrenko}, \citenamefont {Heeres}, \citenamefont {Reinhold},
  \citenamefont {Leghtas}, \citenamefont {Vlastakis}, \citenamefont {Liu},
  \citenamefont {Frunzio}, \citenamefont {Girvin}, \citenamefont {Jiang},
  \citenamefont {Mirrahimi}, \citenamefont {Devoret},\ and\ \citenamefont
  {Schoelkopf}}]{Ofek2016}%
  \BibitemOpen
  \bibfield  {author} {\bibinfo {author} {\bibfnamefont {Nissim}\ \bibnamefont
  {Ofek}}, \bibinfo {author} {\bibfnamefont {Andrei}\ \bibnamefont {Petrenko}},
  \bibinfo {author} {\bibfnamefont {Reinier}\ \bibnamefont {Heeres}}, \bibinfo
  {author} {\bibfnamefont {Philip}\ \bibnamefont {Reinhold}}, \bibinfo {author}
  {\bibfnamefont {Zaki}\ \bibnamefont {Leghtas}}, \bibinfo {author}
  {\bibfnamefont {Brian}\ \bibnamefont {Vlastakis}}, \bibinfo {author}
  {\bibfnamefont {Yehan}\ \bibnamefont {Liu}}, \bibinfo {author} {\bibfnamefont
  {Luigi}\ \bibnamefont {Frunzio}}, \bibinfo {author} {\bibfnamefont {S.~M.}\
  \bibnamefont {Girvin}}, \bibinfo {author} {\bibfnamefont {L.}~\bibnamefont
  {Jiang}}, \bibinfo {author} {\bibfnamefont {Mazyar}\ \bibnamefont
  {Mirrahimi}}, \bibinfo {author} {\bibfnamefont {M.~H.}\ \bibnamefont
  {Devoret}}, \ and\ \bibinfo {author} {\bibfnamefont {R.~J.}\ \bibnamefont
  {Schoelkopf}},\ }\bibfield  {title} {\enquote {\bibinfo {title} {Extending
  the lifetime of a quantum bit with error correction in superconducting
  circuits},}\ }\href {http://dx.doi.org/10.1038/nature18949} {\bibfield
  {journal} {\bibinfo  {journal} {Nature}\ }\textbf {\bibinfo {volume} {536}},\
  \bibinfo {pages} {441--445} (\bibinfo {year} {2016})}\BibitemShut {NoStop}%
\bibitem [{\citenamefont {Rist\`e}\ \emph {et~al.}(2013)\citenamefont
  {Rist\`e}, \citenamefont {Dukalski}, \citenamefont {Watson}, \citenamefont
  {de~Lange}, \citenamefont {Tiggelman}, \citenamefont {Blanter}, \citenamefont
  {Lehnert}, \citenamefont {Schouten},\ and\ \citenamefont
  {DiCarlo}}]{Riste2013}%
  \BibitemOpen
  \bibfield  {author} {\bibinfo {author} {\bibfnamefont {D.}~\bibnamefont
  {Rist\`e}}, \bibinfo {author} {\bibfnamefont {M.}~\bibnamefont {Dukalski}},
  \bibinfo {author} {\bibfnamefont {C.~A.}\ \bibnamefont {Watson}}, \bibinfo
  {author} {\bibfnamefont {G.}~\bibnamefont {de~Lange}}, \bibinfo {author}
  {\bibfnamefont {M.~J.}\ \bibnamefont {Tiggelman}}, \bibinfo {author}
  {\bibfnamefont {Y.~M.}\ \bibnamefont {Blanter}}, \bibinfo {author}
  {\bibfnamefont {K.~W.}\ \bibnamefont {Lehnert}}, \bibinfo {author}
  {\bibfnamefont {R.~N.}\ \bibnamefont {Schouten}}, \ and\ \bibinfo {author}
  {\bibfnamefont {L.}~\bibnamefont {DiCarlo}},\ }\bibfield  {title} {\enquote
  {\bibinfo {title} {Deterministic entanglement of superconducting qubits by
  parity measurement and feedback},}\ }\href {\doibase 10.1038/nature12513}
  {\bibfield  {journal} {\bibinfo  {journal} {Nature}\ }\textbf {\bibinfo
  {volume} {502}},\ \bibinfo {pages} {350--354} (\bibinfo {year}
  {2013})}\BibitemShut {NoStop}%
\bibitem [{\citenamefont {Negnevitsky}\ \emph {et~al.}(2018)\citenamefont
  {Negnevitsky}, \citenamefont {Marinelli}, \citenamefont {Mehta},
  \citenamefont {Lo}, \citenamefont {Flühmann},\ and\ \citenamefont
  {Home}}]{Negnevitsky2018}%
  \BibitemOpen
  \bibfield  {author} {\bibinfo {author} {\bibfnamefont {V.}~\bibnamefont
  {Negnevitsky}}, \bibinfo {author} {\bibfnamefont {M.}~\bibnamefont
  {Marinelli}}, \bibinfo {author} {\bibfnamefont {K.~K.}\ \bibnamefont
  {Mehta}}, \bibinfo {author} {\bibfnamefont {H.-Y.}\ \bibnamefont {Lo}},
  \bibinfo {author} {\bibfnamefont {C.}~\bibnamefont {Flühmann}}, \ and\
  \bibinfo {author} {\bibfnamefont {J.~P.}\ \bibnamefont {Home}},\ }\bibfield
  {title} {\enquote {\bibinfo {title} {Repeated multi-qubit readout and
  feedback with a mixed-species trapped-ion register},}\ }\href
  {https://doi.org/10.1038/s41586-018-0668-z} {\bibfield  {journal} {\bibinfo
  {journal} {Nature}\ }\textbf {\bibinfo {volume} {563}},\ \bibinfo {pages}
  {527--531} (\bibinfo {year} {2018})}\BibitemShut {NoStop}%
\bibitem [{\citenamefont {Andersen}\ \emph {et~al.}(2019)\citenamefont
  {Andersen}, \citenamefont {Remm}, \citenamefont {Balasiu}, \citenamefont
  {Krinner}, \citenamefont {Heinsoo}, \citenamefont {Besse}, \citenamefont
  {Gabureac}, \citenamefont {Wallraff},\ and\ \citenamefont
  {Eichler}}]{Andersen2019}%
  \BibitemOpen
  \bibfield  {author} {\bibinfo {author} {\bibfnamefont {Christian~Kraglund}\
  \bibnamefont {Andersen}}, \bibinfo {author} {\bibfnamefont {Ants}\
  \bibnamefont {Remm}}, \bibinfo {author} {\bibfnamefont {Stefania}\
  \bibnamefont {Balasiu}}, \bibinfo {author} {\bibfnamefont {Sebastian}\
  \bibnamefont {Krinner}}, \bibinfo {author} {\bibfnamefont {Johannes}\
  \bibnamefont {Heinsoo}}, \bibinfo {author} {\bibfnamefont {Jean-Claude}\
  \bibnamefont {Besse}}, \bibinfo {author} {\bibfnamefont {Mihai}\ \bibnamefont
  {Gabureac}}, \bibinfo {author} {\bibfnamefont {Andreas}\ \bibnamefont
  {Wallraff}}, \ and\ \bibinfo {author} {\bibfnamefont {Christopher}\
  \bibnamefont {Eichler}},\ }\bibfield  {title} {\enquote {\bibinfo {title}
  {Entanglement stabilization using ancilla-based parity detection and
  real-time feedback in superconducting circuits},}\ }\href
  {https://doi.org/10.1038/s41534-019-0185-4} {\bibfield  {journal} {\bibinfo
  {journal} {npj Quantum Information}\ }\textbf {\bibinfo {volume} {5}},\
  \bibinfo {pages} {69} (\bibinfo {year} {2019})},\ \Eprint
  {http://arxiv.org/abs/1902.06946} {arXiv:1902.06946 [quant-ph]} \BibitemShut
  {NoStop}%
\bibitem [{\citenamefont {Bultink}\ \emph {et~al.}(2019)\citenamefont
  {Bultink}, \citenamefont {OBrien}, \citenamefont {Vollmer}, \citenamefont
  {Muthusubramanian}, \citenamefont {Beekman}, \citenamefont {Rol},
  \citenamefont {Fu}, \citenamefont {Tarasinski}, \citenamefont {Ostroukh},
  \citenamefont {Varbanov}, \citenamefont {Bruno},\ and\ \citenamefont
  {DiCarlo}}]{Bultink2019}%
  \BibitemOpen
  \bibfield  {author} {\bibinfo {author} {\bibfnamefont {C.~C.}\ \bibnamefont
  {Bultink}}, \bibinfo {author} {\bibfnamefont {T.~E.}\ \bibnamefont {OBrien}},
  \bibinfo {author} {\bibfnamefont {R.}~\bibnamefont {Vollmer}}, \bibinfo
  {author} {\bibfnamefont {N.}~\bibnamefont {Muthusubramanian}}, \bibinfo
  {author} {\bibfnamefont {M.~W.}\ \bibnamefont {Beekman}}, \bibinfo {author}
  {\bibfnamefont {M.~A.}\ \bibnamefont {Rol}}, \bibinfo {author} {\bibfnamefont
  {X.}~\bibnamefont {Fu}}, \bibinfo {author} {\bibfnamefont {B.}~\bibnamefont
  {Tarasinski}}, \bibinfo {author} {\bibfnamefont {V.}~\bibnamefont
  {Ostroukh}}, \bibinfo {author} {\bibfnamefont {B.}~\bibnamefont {Varbanov}},
  \bibinfo {author} {\bibfnamefont {A.}~\bibnamefont {Bruno}}, \ and\ \bibinfo
  {author} {\bibfnamefont {L.}~\bibnamefont {DiCarlo}},\ }\bibfield  {title}
  {\enquote {\bibinfo {title} {Protecting quantum entanglement from qubit
  errors and leakage via repetitive parity measurements},}\ }\href
  {https://arxiv.org/abs/1905.12731} {\bibfield  {journal} {\bibinfo  {journal}
  {arXiv:1905.12731}\ } (\bibinfo {year} {2019})}\BibitemShut {NoStop}%
\bibitem [{\citenamefont {Nigg}\ \emph {et~al.}(2014)\citenamefont {Nigg},
  \citenamefont {M\"uller}, \citenamefont {Martinez}, \citenamefont
  {Schindler}, \citenamefont {Hennrich}, \citenamefont {Monz}, \citenamefont
  {Martin-Delgado},\ and\ \citenamefont {Blatt}}]{Nigg2014b}%
  \BibitemOpen
  \bibfield  {author} {\bibinfo {author} {\bibfnamefont {D.}~\bibnamefont
  {Nigg}}, \bibinfo {author} {\bibfnamefont {M.}~\bibnamefont {M\"uller}},
  \bibinfo {author} {\bibfnamefont {E.~A.}\ \bibnamefont {Martinez}}, \bibinfo
  {author} {\bibfnamefont {P.}~\bibnamefont {Schindler}}, \bibinfo {author}
  {\bibfnamefont {M.}~\bibnamefont {Hennrich}}, \bibinfo {author}
  {\bibfnamefont {T.}~\bibnamefont {Monz}}, \bibinfo {author} {\bibfnamefont
  {M.~A.}\ \bibnamefont {Martin-Delgado}}, \ and\ \bibinfo {author}
  {\bibfnamefont {R.}~\bibnamefont {Blatt}},\ }\bibfield  {title} {\enquote
  {\bibinfo {title} {Quantum computations on a topologically encoded qubit},}\
  }\href {\doibase 10.1126/science.1253742} {\bibfield  {journal} {\bibinfo
  {journal} {Science}\ }\textbf {\bibinfo {volume} {345}},\ \bibinfo {pages}
  {302--305} (\bibinfo {year} {2014})}\BibitemShut {NoStop}%
\bibitem [{\citenamefont {Gong}\ \emph {et~al.}(2019)\citenamefont {Gong},
  \citenamefont {Yuan}, \citenamefont {Wang}, \citenamefont {Wu}, \citenamefont
  {Zhao}, \citenamefont {Zha}, \citenamefont {Li}, \citenamefont {Zhang},
  \citenamefont {Zhao}, \citenamefont {Liu}, \citenamefont {Liang},
  \citenamefont {Lin}, \citenamefont {Xu}, \citenamefont {Deng}, \citenamefont
  {Rong}, \citenamefont {Lu}, \citenamefont {Benjamin}, \citenamefont {Peng},
  \citenamefont {Ma}, \citenamefont {Chen}, \citenamefont {Zhu}, \citenamefont
  {Pan}, \citenamefont {Peng}, \citenamefont {Ma}, \citenamefont {Chen},
  \citenamefont {Zhu},\ and\ \citenamefont {Pan}}]{Gong2019a}%
  \BibitemOpen
  \bibfield  {author} {\bibinfo {author} {\bibfnamefont {M.}~\bibnamefont
  {Gong}}, \bibinfo {author} {\bibfnamefont {X.}~\bibnamefont {Yuan}}, \bibinfo
  {author} {\bibfnamefont {S.}~\bibnamefont {Wang}}, \bibinfo {author}
  {\bibfnamefont {Y.}~\bibnamefont {Wu}}, \bibinfo {author} {\bibfnamefont
  {Y.}~\bibnamefont {Zhao}}, \bibinfo {author} {\bibfnamefont {C.}~\bibnamefont
  {Zha}}, \bibinfo {author} {\bibfnamefont {S.}~\bibnamefont {Li}}, \bibinfo
  {author} {\bibfnamefont {Z.}~\bibnamefont {Zhang}}, \bibinfo {author}
  {\bibfnamefont {Q.}~\bibnamefont {Zhao}}, \bibinfo {author} {\bibfnamefont
  {Y.}~\bibnamefont {Liu}}, \bibinfo {author} {\bibfnamefont {F.}~\bibnamefont
  {Liang}}, \bibinfo {author} {\bibfnamefont {J.}~\bibnamefont {Lin}}, \bibinfo
  {author} {\bibfnamefont {Y.}~\bibnamefont {Xu}}, \bibinfo {author}
  {\bibfnamefont {H.}~\bibnamefont {Deng}}, \bibinfo {author} {\bibfnamefont
  {H.}~\bibnamefont {Rong}}, \bibinfo {author} {\bibfnamefont {H.}~\bibnamefont
  {Lu}}, \bibinfo {author} {\bibfnamefont {S.~C.}\ \bibnamefont {Benjamin}},
  \bibinfo {author} {\bibfnamefont {C.}~\bibnamefont {Peng}}, \bibinfo {author}
  {\bibfnamefont {X.}~\bibnamefont {Ma}}, \bibinfo {author} {\bibfnamefont
  {Y.}~\bibnamefont {Chen}}, \bibinfo {author} {\bibfnamefont {X.}~\bibnamefont
  {Zhu}}, \bibinfo {author} {\bibfnamefont {S.~C.}\ \bibnamefont {Pan},
  \bibfnamefont {J.enjamin}}, \bibinfo {author} {\bibfnamefont
  {C.}~\bibnamefont {Peng}}, \bibinfo {author} {\bibfnamefont {X.}~\bibnamefont
  {Ma}}, \bibinfo {author} {\bibfnamefont {Y.}~\bibnamefont {Chen}}, \bibinfo
  {author} {\bibfnamefont {X.}~\bibnamefont {Zhu}}, \ and\ \bibinfo {author}
  {\bibfnamefont {J.}~\bibnamefont {Pan}},\ }\bibfield  {title} {\enquote
  {\bibinfo {title} {Experimental verification of five-qubit quantum error
  correction with superconducting qubits},}\ }\href
  {https://arxiv.org/abs/1907.04507} {\bibfield  {journal} {\bibinfo  {journal}
  {arXiv:1907.04507}\ } (\bibinfo {year} {2019})}\BibitemShut {NoStop}%
\bibitem [{\citenamefont {Takita}\ \emph {et~al.}(2017)\citenamefont {Takita},
  \citenamefont {Cross}, \citenamefont {C\'orcoles}, \citenamefont {Chow},\
  and\ \citenamefont {Gambetta}}]{Takita2017}%
  \BibitemOpen
  \bibfield  {author} {\bibinfo {author} {\bibfnamefont {Maika}\ \bibnamefont
  {Takita}}, \bibinfo {author} {\bibfnamefont {Andrew~W.}\ \bibnamefont
  {Cross}}, \bibinfo {author} {\bibfnamefont {A.~D.}\ \bibnamefont
  {C\'orcoles}}, \bibinfo {author} {\bibfnamefont {Jerry~M.}\ \bibnamefont
  {Chow}}, \ and\ \bibinfo {author} {\bibfnamefont {Jay~M.}\ \bibnamefont
  {Gambetta}},\ }\bibfield  {title} {\enquote {\bibinfo {title} {Experimental
  demonstration of fault-tolerant state preparation with superconducting
  qubits},}\ }\href {\doibase 10.1103/PhysRevLett.119.180501} {\bibfield
  {journal} {\bibinfo  {journal} {Phys. Rev. Lett.}\ }\textbf {\bibinfo
  {volume} {119}},\ \bibinfo {pages} {180501} (\bibinfo {year}
  {2017})}\BibitemShut {NoStop}%
\bibitem [{\citenamefont {Hu}\ \emph {et~al.}(2019)\citenamefont {Hu},
  \citenamefont {Ma}, \citenamefont {Cai}, \citenamefont {Mu}, \citenamefont
  {Xu}, \citenamefont {Wang}, \citenamefont {Wu}, \citenamefont {Wang},
  \citenamefont {Song}, \citenamefont {Zou}, \citenamefont {Girvin},
  \citenamefont {Duan},\ and\ \citenamefont {Sun}}]{Hu2019a}%
  \BibitemOpen
  \bibfield  {author} {\bibinfo {author} {\bibfnamefont {L.}~\bibnamefont
  {Hu}}, \bibinfo {author} {\bibfnamefont {Y.}~\bibnamefont {Ma}}, \bibinfo
  {author} {\bibfnamefont {W.}~\bibnamefont {Cai}}, \bibinfo {author}
  {\bibfnamefont {X.}~\bibnamefont {Mu}}, \bibinfo {author} {\bibfnamefont
  {Y.}~\bibnamefont {Xu}}, \bibinfo {author} {\bibfnamefont {W.}~\bibnamefont
  {Wang}}, \bibinfo {author} {\bibfnamefont {Y.}~\bibnamefont {Wu}}, \bibinfo
  {author} {\bibfnamefont {H.}~\bibnamefont {Wang}}, \bibinfo {author}
  {\bibfnamefont {Y.~P.}\ \bibnamefont {Song}}, \bibinfo {author}
  {\bibfnamefont {C.-L.}\ \bibnamefont {Zou}}, \bibinfo {author} {\bibfnamefont
  {S.~M.}\ \bibnamefont {Girvin}}, \bibinfo {author} {\bibfnamefont {L.-M.}\
  \bibnamefont {Duan}}, \ and\ \bibinfo {author} {\bibfnamefont
  {L.}~\bibnamefont {Sun}},\ }\bibfield  {title} {\enquote {\bibinfo {title}
  {Quantum error correction and universal gate set operation on a binomial
  bosonic logical qubit},}\ }\href {https://doi.org/10.1038/s41567-018-0414-3}
  {\bibfield  {journal} {\bibinfo  {journal} {Nature Physics}\ } (\bibinfo
  {year} {2019})}\BibitemShut {NoStop}%
\bibitem [{\citenamefont {Campagne-Ibarcq}\ \emph {et~al.}(2019)\citenamefont
  {Campagne-Ibarcq}, \citenamefont {Eickbusch}, \citenamefont {Touzard},
  \citenamefont {Zalys-Geller}, \citenamefont {Frattini}, \citenamefont
  {Sivak}, \citenamefont {Reinhold}, \citenamefont {Puri}, \citenamefont
  {Shankar}, \citenamefont {Schoelkopf}, \citenamefont {Frunzio}, \citenamefont
  {Mirrahimi},\ and\ \citenamefont {Devoret}}]{Campagne-Ibarcq2019}%
  \BibitemOpen
  \bibfield  {author} {\bibinfo {author} {\bibfnamefont {P.}~\bibnamefont
  {Campagne-Ibarcq}}, \bibinfo {author} {\bibfnamefont {A.}~\bibnamefont
  {Eickbusch}}, \bibinfo {author} {\bibfnamefont {S.}~\bibnamefont {Touzard}},
  \bibinfo {author} {\bibfnamefont {E.}~\bibnamefont {Zalys-Geller}}, \bibinfo
  {author} {\bibfnamefont {N.~E.}\ \bibnamefont {Frattini}}, \bibinfo {author}
  {\bibfnamefont {V.~V.}\ \bibnamefont {Sivak}}, \bibinfo {author}
  {\bibfnamefont {P.}~\bibnamefont {Reinhold}}, \bibinfo {author}
  {\bibfnamefont {S.}~\bibnamefont {Puri}}, \bibinfo {author} {\bibfnamefont
  {S.}~\bibnamefont {Shankar}}, \bibinfo {author} {\bibfnamefont {R.~J.}\
  \bibnamefont {Schoelkopf}}, \bibinfo {author} {\bibfnamefont
  {L.}~\bibnamefont {Frunzio}}, \bibinfo {author} {\bibfnamefont
  {M.}~\bibnamefont {Mirrahimi}}, \ and\ \bibinfo {author} {\bibfnamefont
  {M.~H.}\ \bibnamefont {Devoret}},\ }\bibfield  {title} {\enquote {\bibinfo
  {title} {A stabilized logical quantum bit encoded in grid states of a
  superconducting cavity},}\ }\href {https://arxiv.org/abs/1907.12487}
  {\bibfield  {journal} {\bibinfo  {journal} {arXiv:1907.12487}\ } (\bibinfo
  {year} {2019})}\BibitemShut {NoStop}%
\bibitem [{\citenamefont {Fowler}\ \emph {et~al.}(2012)\citenamefont {Fowler},
  \citenamefont {Mariantoni}, \citenamefont {Martinis},\ and\ \citenamefont
  {Cleland}}]{Fowler2012}%
  \BibitemOpen
  \bibfield  {author} {\bibinfo {author} {\bibfnamefont {Austin~G.}\
  \bibnamefont {Fowler}}, \bibinfo {author} {\bibfnamefont {Matteo}\
  \bibnamefont {Mariantoni}}, \bibinfo {author} {\bibfnamefont {John~M.}\
  \bibnamefont {Martinis}}, \ and\ \bibinfo {author} {\bibfnamefont
  {Andrew~N.}\ \bibnamefont {Cleland}},\ }\bibfield  {title} {\enquote
  {\bibinfo {title} {Surface codes: Towards practical large-scale quantum
  computation},}\ }\href {\doibase 10.1103/PhysRevA.86.032324} {\bibfield
  {journal} {\bibinfo  {journal} {Phys. Rev. A}\ }\textbf {\bibinfo {volume}
  {86}},\ \bibinfo {pages} {032324} (\bibinfo {year} {2012})}\BibitemShut
  {NoStop}%
\bibitem [{\citenamefont {Lidar}\ and\ \citenamefont {Brun}(2013)}]{Lidar2013}%
  \BibitemOpen
  \bibfield  {author} {\bibinfo {author} {\bibfnamefont {Daniel~A.}\
  \bibnamefont {Lidar}}\ and\ \bibinfo {author} {\bibfnamefont {Todd~A.}\
  \bibnamefont {Brun}},\ }\href@noop {} {\emph {\bibinfo {title} {Quantum Error
  Correction}}}\ (\bibinfo  {publisher} {Cambridge University Press},\ \bibinfo
  {year} {2013})\BibitemShut {NoStop}%
\bibitem [{\citenamefont {Terhal}(2015)}]{Terhal2015n}%
  \BibitemOpen
  \bibfield  {author} {\bibinfo {author} {\bibfnamefont {Barbara~M.}\
  \bibnamefont {Terhal}},\ }\bibfield  {title} {\enquote {\bibinfo {title}
  {Quantum error correction for quantum memories},}\ }\href {\doibase
  10.1103/RevModPhys.87.307} {\bibfield  {journal} {\bibinfo  {journal} {Rev.
  Mod. Phys.}\ }\textbf {\bibinfo {volume} {87}},\ \bibinfo {pages} {307--346}
  (\bibinfo {year} {2015})}\BibitemShut {NoStop}%
\bibitem [{\citenamefont {Groen}\ \emph {et~al.}(2013)\citenamefont {Groen},
  \citenamefont {Rista}, \citenamefont {Tornberg}, \citenamefont {Cramer},
  \citenamefont {de~Groot}, \citenamefont {Picot}, \citenamefont {Johansson},\
  and\ \citenamefont {DiCarlo}}]{Groen2013}%
  \BibitemOpen
  \bibfield  {author} {\bibinfo {author} {\bibfnamefont {J.~P.}\ \bibnamefont
  {Groen}}, \bibinfo {author} {\bibfnamefont {D.}~\bibnamefont {Rista}},
  \bibinfo {author} {\bibfnamefont {L.}~\bibnamefont {Tornberg}}, \bibinfo
  {author} {\bibfnamefont {J.}~\bibnamefont {Cramer}}, \bibinfo {author}
  {\bibfnamefont {P.~C.}\ \bibnamefont {de~Groot}}, \bibinfo {author}
  {\bibfnamefont {T.}~\bibnamefont {Picot}}, \bibinfo {author} {\bibfnamefont
  {G.}~\bibnamefont {Johansson}}, \ and\ \bibinfo {author} {\bibfnamefont
  {L.}~\bibnamefont {DiCarlo}},\ }\bibfield  {title} {\enquote {\bibinfo
  {title} {Partial-measurement backaction and nonclassical weak values in a
  superconducting circuit},}\ }\href
  {http://link.aps.org/doi/10.1103/PhysRevLett.111.090506} {\bibfield
  {journal} {\bibinfo  {journal} {Phys. Rev. Lett.}\ }\textbf {\bibinfo
  {volume} {111}},\ \bibinfo {pages} {090506} (\bibinfo {year}
  {2013})}\BibitemShut {NoStop}%
\bibitem [{\citenamefont {Schmitt}\ \emph {et~al.}(2014)\citenamefont
  {Schmitt}, \citenamefont {Zhou}, \citenamefont {Juliusson}, \citenamefont
  {Royer}, \citenamefont {Blais}, \citenamefont {Bertet}, \citenamefont
  {Vion},\ and\ \citenamefont {Esteve}}]{Schmitt2014a}%
  \BibitemOpen
  \bibfield  {author} {\bibinfo {author} {\bibfnamefont {V.}~\bibnamefont
  {Schmitt}}, \bibinfo {author} {\bibfnamefont {X.}~\bibnamefont {Zhou}},
  \bibinfo {author} {\bibfnamefont {K.}~\bibnamefont {Juliusson}}, \bibinfo
  {author} {\bibfnamefont {B.}~\bibnamefont {Royer}}, \bibinfo {author}
  {\bibfnamefont {A.}~\bibnamefont {Blais}}, \bibinfo {author} {\bibfnamefont
  {P.}~\bibnamefont {Bertet}}, \bibinfo {author} {\bibfnamefont
  {D.}~\bibnamefont {Vion}}, \ and\ \bibinfo {author} {\bibfnamefont
  {D.}~\bibnamefont {Esteve}},\ }\bibfield  {title} {\enquote {\bibinfo {title}
  {Multiplexed readout of transmon qubits with {Josephson} bifurcation
  amplifiers},}\ }\href {\doibase 10.1103/PhysRevA.90.062333} {\bibfield
  {journal} {\bibinfo  {journal} {Phys. Rev. A}\ }\textbf {\bibinfo {volume}
  {90}},\ \bibinfo {pages} {062333} (\bibinfo {year} {2014})}\BibitemShut
  {NoStop}%
\bibitem [{\citenamefont {Jeffrey}\ \emph {et~al.}(2014)\citenamefont
  {Jeffrey}, \citenamefont {Sank}, \citenamefont {Mutus}, \citenamefont
  {White}, \citenamefont {Kelly}, \citenamefont {Barends}, \citenamefont
  {Chen}, \citenamefont {Chen}, \citenamefont {Chiaro}, \citenamefont
  {Dunsworth}, \citenamefont {Megrant}, \citenamefont {O'Malley}, \citenamefont
  {Neill}, \citenamefont {Roushan}, \citenamefont {Vainsencher}, \citenamefont
  {Wenner}, \citenamefont {Cleland},\ and\ \citenamefont
  {Martinis}}]{Jeffrey2014}%
  \BibitemOpen
  \bibfield  {author} {\bibinfo {author} {\bibfnamefont {E.}~\bibnamefont
  {Jeffrey}}, \bibinfo {author} {\bibfnamefont {D.}~\bibnamefont {Sank}},
  \bibinfo {author} {\bibfnamefont {J.~Y.}\ \bibnamefont {Mutus}}, \bibinfo
  {author} {\bibfnamefont {T.~C.}\ \bibnamefont {White}}, \bibinfo {author}
  {\bibfnamefont {J.}~\bibnamefont {Kelly}}, \bibinfo {author} {\bibfnamefont
  {R.}~\bibnamefont {Barends}}, \bibinfo {author} {\bibfnamefont
  {Y.}~\bibnamefont {Chen}}, \bibinfo {author} {\bibfnamefont {Z.}~\bibnamefont
  {Chen}}, \bibinfo {author} {\bibfnamefont {B.}~\bibnamefont {Chiaro}},
  \bibinfo {author} {\bibfnamefont {A.}~\bibnamefont {Dunsworth}}, \bibinfo
  {author} {\bibfnamefont {A.}~\bibnamefont {Megrant}}, \bibinfo {author}
  {\bibfnamefont {P.~J.~J.}\ \bibnamefont {O'Malley}}, \bibinfo {author}
  {\bibfnamefont {C.}~\bibnamefont {Neill}}, \bibinfo {author} {\bibfnamefont
  {P.}~\bibnamefont {Roushan}}, \bibinfo {author} {\bibfnamefont
  {A.}~\bibnamefont {Vainsencher}}, \bibinfo {author} {\bibfnamefont
  {J.}~\bibnamefont {Wenner}}, \bibinfo {author} {\bibfnamefont {A.~N.}\
  \bibnamefont {Cleland}}, \ and\ \bibinfo {author} {\bibfnamefont {J.~M.}\
  \bibnamefont {Martinis}},\ }\bibfield  {title} {\enquote {\bibinfo {title}
  {Fast accurate state measurement with superconducting qubits},}\ }\href
  {\doibase 10.1103/PhysRevLett.112.190504} {\bibfield  {journal} {\bibinfo
  {journal} {Phys. Rev. Lett.}\ }\textbf {\bibinfo {volume} {112}},\ \bibinfo
  {pages} {190504} (\bibinfo {year} {2014})}\BibitemShut {NoStop}%
\bibitem [{\citenamefont {Heinsoo}\ \emph {et~al.}(2018)\citenamefont
  {Heinsoo}, \citenamefont {Andersen}, \citenamefont {Remm}, \citenamefont
  {Krinner}, \citenamefont {Walter}, \citenamefont {Salath\'{e}}, \citenamefont
  {Gasparinetti}, \citenamefont {Besse}, \citenamefont {Poto\v{c}nik},
  \citenamefont {Wallraff},\ and\ \citenamefont {Eichler}}]{Heinsoo2018}%
  \BibitemOpen
  \bibfield  {author} {\bibinfo {author} {\bibfnamefont {Johannes}\
  \bibnamefont {Heinsoo}}, \bibinfo {author} {\bibfnamefont
  {Christian~Kraglund}\ \bibnamefont {Andersen}}, \bibinfo {author}
  {\bibfnamefont {Ants}\ \bibnamefont {Remm}}, \bibinfo {author} {\bibfnamefont
  {Sebastian}\ \bibnamefont {Krinner}}, \bibinfo {author} {\bibfnamefont
  {Theodore}\ \bibnamefont {Walter}}, \bibinfo {author} {\bibfnamefont {Yves}\
  \bibnamefont {Salath\'{e}}}, \bibinfo {author} {\bibfnamefont {Simone}\
  \bibnamefont {Gasparinetti}}, \bibinfo {author} {\bibfnamefont {Jean-Claude}\
  \bibnamefont {Besse}}, \bibinfo {author} {\bibfnamefont {Anton}\ \bibnamefont
  {Poto\v{c}nik}}, \bibinfo {author} {\bibfnamefont {Andreas}\ \bibnamefont
  {Wallraff}}, \ and\ \bibinfo {author} {\bibfnamefont {Christopher}\
  \bibnamefont {Eichler}},\ }\bibfield  {title} {\enquote {\bibinfo {title}
  {Rapid high-fidelity multiplexed readout of superconducting qubits},}\ }\href
  {\doibase 10.1103/PhysRevApplied.10.034040} {\bibfield  {journal} {\bibinfo
  {journal} {Phys. Rev. Applied}\ }\textbf {\bibinfo {volume} {10}},\ \bibinfo
  {pages} {034040} (\bibinfo {year} {2018})}\BibitemShut {NoStop}%
\bibitem [{\citenamefont {Barends}\ \emph {et~al.}(2014)\citenamefont
  {Barends}, \citenamefont {Kelly}, \citenamefont {Megrant}, \citenamefont
  {Veitia}, \citenamefont {Sank}, \citenamefont {Jeffrey}, \citenamefont
  {White}, \citenamefont {Mutus}, \citenamefont {Fowler}, \citenamefont
  {Campbell}, \citenamefont {Chen}, \citenamefont {Chen}, \citenamefont
  {Chiaro}, \citenamefont {Dunsworth}, \citenamefont {Neill}, \citenamefont
  {{O\'Malley}}, \citenamefont {Roushan}, \citenamefont {Vainsencher},
  \citenamefont {Wenner}, \citenamefont {Korotkov}, \citenamefont {Cleland},\
  and\ \citenamefont {Martinis}}]{Barends2014}%
  \BibitemOpen
  \bibfield  {author} {\bibinfo {author} {\bibfnamefont {R.}~\bibnamefont
  {Barends}}, \bibinfo {author} {\bibfnamefont {J.}~\bibnamefont {Kelly}},
  \bibinfo {author} {\bibfnamefont {A.}~\bibnamefont {Megrant}}, \bibinfo
  {author} {\bibfnamefont {A.}~\bibnamefont {Veitia}}, \bibinfo {author}
  {\bibfnamefont {D.}~\bibnamefont {Sank}}, \bibinfo {author} {\bibfnamefont
  {E.}~\bibnamefont {Jeffrey}}, \bibinfo {author} {\bibfnamefont {T.~C.}\
  \bibnamefont {White}}, \bibinfo {author} {\bibfnamefont {J.}~\bibnamefont
  {Mutus}}, \bibinfo {author} {\bibfnamefont {A.~G.}\ \bibnamefont {Fowler}},
  \bibinfo {author} {\bibfnamefont {B.}~\bibnamefont {Campbell}}, \bibinfo
  {author} {\bibfnamefont {Y.}~\bibnamefont {Chen}}, \bibinfo {author}
  {\bibfnamefont {Z.}~\bibnamefont {Chen}}, \bibinfo {author} {\bibfnamefont
  {B.}~\bibnamefont {Chiaro}}, \bibinfo {author} {\bibfnamefont
  {A.}~\bibnamefont {Dunsworth}}, \bibinfo {author} {\bibfnamefont
  {C.}~\bibnamefont {Neill}}, \bibinfo {author} {\bibfnamefont
  {P.}~\bibnamefont {{O\'Malley}}}, \bibinfo {author} {\bibfnamefont
  {P.}~\bibnamefont {Roushan}}, \bibinfo {author} {\bibfnamefont
  {A.}~\bibnamefont {Vainsencher}}, \bibinfo {author} {\bibfnamefont
  {J.}~\bibnamefont {Wenner}}, \bibinfo {author} {\bibfnamefont {A.~N.}\
  \bibnamefont {Korotkov}}, \bibinfo {author} {\bibfnamefont {A.~N.}\
  \bibnamefont {Cleland}}, \ and\ \bibinfo {author} {\bibfnamefont {John~M.}\
  \bibnamefont {Martinis}},\ }\bibfield  {title} {\enquote {\bibinfo {title}
  {Superconducting quantum circuits at the surface code threshold for fault
  tolerance},}\ }\href {\doibase 10.1038/nature13171} {\bibfield  {journal}
  {\bibinfo  {journal} {Nature}\ }\textbf {\bibinfo {volume} {508}},\ \bibinfo
  {pages} {500--503} (\bibinfo {year} {2014})}\BibitemShut {NoStop}%
\bibitem [{\citenamefont {Rol}\ \emph {et~al.}(2019)\citenamefont {Rol},
  \citenamefont {Battistel}, \citenamefont {Malinowski}, \citenamefont
  {Bultink}, \citenamefont {Tarasinski}, \citenamefont {Vollmer}, \citenamefont
  {Haider}, \citenamefont {Muthusubramanian}, \citenamefont {Bruno},
  \citenamefont {Terhal},\ and\ \citenamefont {DiCarlo}}]{Rol2019}%
  \BibitemOpen
  \bibfield  {author} {\bibinfo {author} {\bibfnamefont {M.~A.}\ \bibnamefont
  {Rol}}, \bibinfo {author} {\bibfnamefont {F.}~\bibnamefont {Battistel}},
  \bibinfo {author} {\bibfnamefont {F.~K.}\ \bibnamefont {Malinowski}},
  \bibinfo {author} {\bibfnamefont {C.~C.}\ \bibnamefont {Bultink}}, \bibinfo
  {author} {\bibfnamefont {B.~M.}\ \bibnamefont {Tarasinski}}, \bibinfo
  {author} {\bibfnamefont {R.}~\bibnamefont {Vollmer}}, \bibinfo {author}
  {\bibfnamefont {N.}~\bibnamefont {Haider}}, \bibinfo {author} {\bibfnamefont
  {N.}~\bibnamefont {Muthusubramanian}}, \bibinfo {author} {\bibfnamefont
  {A.}~\bibnamefont {Bruno}}, \bibinfo {author} {\bibfnamefont {B.~M.}\
  \bibnamefont {Terhal}}, \ and\ \bibinfo {author} {\bibfnamefont
  {L.}~\bibnamefont {DiCarlo}},\ }\bibfield  {title} {\enquote {\bibinfo
  {title} {Fast, high-fidelity conditional-phase gate exploiting leakage
  interference in weakly anharmonic superconducting qubits},}\ }\href {\doibase
  10.1103/PhysRevLett.123.120502} {\bibfield  {journal} {\bibinfo  {journal}
  {Phys. Rev. Lett.}\ }\textbf {\bibinfo {volume} {123}},\ \bibinfo {pages}
  {120502} (\bibinfo {year} {2019})}\BibitemShut {NoStop}%
\bibitem [{\citenamefont {Versluis}\ \emph {et~al.}(2017)\citenamefont
  {Versluis}, \citenamefont {Poletto}, \citenamefont {Khammassi}, \citenamefont
  {Tarasinski}, \citenamefont {Haider}, \citenamefont {Michalak}, \citenamefont
  {Bruno}, \citenamefont {Bertels},\ and\ \citenamefont
  {DiCarlo}}]{Versluis2017}%
  \BibitemOpen
  \bibfield  {author} {\bibinfo {author} {\bibfnamefont {R.}~\bibnamefont
  {Versluis}}, \bibinfo {author} {\bibfnamefont {S.}~\bibnamefont {Poletto}},
  \bibinfo {author} {\bibfnamefont {N.}~\bibnamefont {Khammassi}}, \bibinfo
  {author} {\bibfnamefont {B.}~\bibnamefont {Tarasinski}}, \bibinfo {author}
  {\bibfnamefont {N.}~\bibnamefont {Haider}}, \bibinfo {author} {\bibfnamefont
  {D.~J.}\ \bibnamefont {Michalak}}, \bibinfo {author} {\bibfnamefont
  {A.}~\bibnamefont {Bruno}}, \bibinfo {author} {\bibfnamefont
  {K.}~\bibnamefont {Bertels}}, \ and\ \bibinfo {author} {\bibfnamefont
  {L.}~\bibnamefont {DiCarlo}},\ }\bibfield  {title} {\enquote {\bibinfo
  {title} {Scalable quantum circuit and control for a superconducting surface
  code},}\ }\href {\doibase 10.1103/PhysRevApplied.8.034021} {\bibfield
  {journal} {\bibinfo  {journal} {Phys. Rev. Applied}\ }\textbf {\bibinfo
  {volume} {8}},\ \bibinfo {pages} {034021} (\bibinfo {year}
  {2017})}\BibitemShut {NoStop}%
\bibitem [{\citenamefont {Johnson}\ \emph {et~al.}(2012)\citenamefont
  {Johnson}, \citenamefont {Macklin}, \citenamefont {Slichter}, \citenamefont
  {Vijay}, \citenamefont {Weingarten}, \citenamefont {Clarke},\ and\
  \citenamefont {Siddiqi}}]{Johnson2012}%
  \BibitemOpen
  \bibfield  {author} {\bibinfo {author} {\bibfnamefont {J.~E.}\ \bibnamefont
  {Johnson}}, \bibinfo {author} {\bibfnamefont {C.}~\bibnamefont {Macklin}},
  \bibinfo {author} {\bibfnamefont {D.~H.}\ \bibnamefont {Slichter}}, \bibinfo
  {author} {\bibfnamefont {R.}~\bibnamefont {Vijay}}, \bibinfo {author}
  {\bibfnamefont {E.~B.}\ \bibnamefont {Weingarten}}, \bibinfo {author}
  {\bibfnamefont {John}\ \bibnamefont {Clarke}}, \ and\ \bibinfo {author}
  {\bibfnamefont {I.}~\bibnamefont {Siddiqi}},\ }\bibfield  {title} {\enquote
  {\bibinfo {title} {Heralded state preparation in a superconducting qubit},}\
  }\href {\doibase 10.1103/PhysRevLett.109.050506} {\bibfield  {journal}
  {\bibinfo  {journal} {Phys. Rev. Lett.}\ }\textbf {\bibinfo {volume} {109}},\
  \bibinfo {pages} {050506} (\bibinfo {year} {2012})}\BibitemShut {NoStop}%
\bibitem [{\citenamefont {Rist\`e}\ \emph {et~al.}(2012)\citenamefont
  {Rist\`e}, \citenamefont {van Leeuwen}, \citenamefont {Ku}, \citenamefont
  {Lehnert},\ and\ \citenamefont {DiCarlo}}]{Riste2012}%
  \BibitemOpen
  \bibfield  {author} {\bibinfo {author} {\bibfnamefont {D.}~\bibnamefont
  {Rist\`e}}, \bibinfo {author} {\bibfnamefont {J.~G.}\ \bibnamefont {van
  Leeuwen}}, \bibinfo {author} {\bibfnamefont {H.-S.}\ \bibnamefont {Ku}},
  \bibinfo {author} {\bibfnamefont {K.~W.}\ \bibnamefont {Lehnert}}, \ and\
  \bibinfo {author} {\bibfnamefont {L.}~\bibnamefont {DiCarlo}},\ }\bibfield
  {title} {\enquote {\bibinfo {title} {Initialization by measurement of a
  superconducting quantum bit circuit},}\ }\href {\doibase
  10.1103/PhysRevLett.109.050507} {\bibfield  {journal} {\bibinfo  {journal}
  {Phys. Rev. Lett.}\ }\textbf {\bibinfo {volume} {109}},\ \bibinfo {pages}
  {050507} (\bibinfo {year} {2012})}\BibitemShut {NoStop}%
\bibitem [{\citenamefont {Koch}\ \emph {et~al.}(2009)\citenamefont {Koch},
  \citenamefont {Manucharyan}, \citenamefont {Devoret},\ and\ \citenamefont
  {Glazman}}]{Koch2009}%
  \BibitemOpen
  \bibfield  {author} {\bibinfo {author} {\bibfnamefont {J.}~\bibnamefont
  {Koch}}, \bibinfo {author} {\bibfnamefont {V.~E.}\ \bibnamefont
  {Manucharyan}}, \bibinfo {author} {\bibfnamefont {M.~H.}\ \bibnamefont
  {Devoret}}, \ and\ \bibinfo {author} {\bibfnamefont {L.~I.}\ \bibnamefont
  {Glazman}},\ }\href@noop {} {\enquote {\bibinfo {title} {Charging effects in
  the inductively shunted {Josephson} junction},}\ } (\bibinfo {year}
  {2009})\BibitemShut {NoStop}%
\bibitem [{\citenamefont {Reed}\ \emph {et~al.}(2010)\citenamefont {Reed},
  \citenamefont {Johnson}, \citenamefont {Houck}, \citenamefont {DiCarlo},
  \citenamefont {Chow}, \citenamefont {Schuster}, \citenamefont {Frunzio},\
  and\ \citenamefont {Schoelkopf}}]{Reed2010}%
  \BibitemOpen
  \bibfield  {author} {\bibinfo {author} {\bibfnamefont {M.~D.}\ \bibnamefont
  {Reed}}, \bibinfo {author} {\bibfnamefont {B.~R.}\ \bibnamefont {Johnson}},
  \bibinfo {author} {\bibfnamefont {A.~A.}\ \bibnamefont {Houck}}, \bibinfo
  {author} {\bibfnamefont {L.}~\bibnamefont {DiCarlo}}, \bibinfo {author}
  {\bibfnamefont {J.~M.}\ \bibnamefont {Chow}}, \bibinfo {author}
  {\bibfnamefont {D.~I.}\ \bibnamefont {Schuster}}, \bibinfo {author}
  {\bibfnamefont {L.}~\bibnamefont {Frunzio}}, \ and\ \bibinfo {author}
  {\bibfnamefont {R.~J.}\ \bibnamefont {Schoelkopf}},\ }\bibfield  {title}
  {\enquote {\bibinfo {title} {Fast reset and suppressing spontaneous emission
  of a superconducting qubit},}\ }\href {\doibase 10.1063/1.3435463} {\bibfield
   {journal} {\bibinfo  {journal} {Appl. Phys. Lett.}\ }\textbf {\bibinfo
  {volume} {96}},\ \bibinfo {eid} {203110} (\bibinfo {year}
  {2010})}\BibitemShut {NoStop}%
\bibitem [{\citenamefont {Krinner}\ \emph {et~al.}(2019)\citenamefont
  {Krinner}, \citenamefont {Storz}, \citenamefont {Kurpiers}, \citenamefont
  {Magnard}, \citenamefont {Heinsoo}, \citenamefont {Keller}, \citenamefont
  {L{\"u}tolf}, \citenamefont {Eichler},\ and\ \citenamefont
  {Wallraff}}]{Krinner2019}%
  \BibitemOpen
  \bibfield  {author} {\bibinfo {author} {\bibfnamefont {S.}~\bibnamefont
  {Krinner}}, \bibinfo {author} {\bibfnamefont {S.}~\bibnamefont {Storz}},
  \bibinfo {author} {\bibfnamefont {P.}~\bibnamefont {Kurpiers}}, \bibinfo
  {author} {\bibfnamefont {P.}~\bibnamefont {Magnard}}, \bibinfo {author}
  {\bibfnamefont {J.}~\bibnamefont {Heinsoo}}, \bibinfo {author} {\bibfnamefont
  {R.}~\bibnamefont {Keller}}, \bibinfo {author} {\bibfnamefont
  {J.}~\bibnamefont {L{\"u}tolf}}, \bibinfo {author} {\bibfnamefont
  {C.}~\bibnamefont {Eichler}}, \ and\ \bibinfo {author} {\bibfnamefont
  {A.}~\bibnamefont {Wallraff}},\ }\bibfield  {title} {\enquote {\bibinfo
  {title} {Engineering cryogenic setups for 100-qubit scale superconducting
  circuit systems},}\ }\href {\doibase 10.1140/epjqt/s40507-019-0072-0}
  {\bibfield  {journal} {\bibinfo  {journal} {EPJ Quantum Technology}\ }\textbf
  {\bibinfo {volume} {6}},\ \bibinfo {pages} {2} (\bibinfo {year}
  {2019})}\BibitemShut {NoStop}%
\bibitem [{\citenamefont {Takita}\ \emph {et~al.}(2016)\citenamefont {Takita},
  \citenamefont {C\'{o}rcoles}, \citenamefont {Magesan}, \citenamefont {Abdo},
  \citenamefont {Brink}, \citenamefont {Cross}, \citenamefont {Chow},\ and\
  \citenamefont {Gambetta}}]{Takita2016}%
  \BibitemOpen
  \bibfield  {author} {\bibinfo {author} {\bibfnamefont {Maika}\ \bibnamefont
  {Takita}}, \bibinfo {author} {\bibfnamefont {A.D.}\ \bibnamefont
  {C\'{o}rcoles}}, \bibinfo {author} {\bibfnamefont {Easwar}\ \bibnamefont
  {Magesan}}, \bibinfo {author} {\bibfnamefont {Baleegh}\ \bibnamefont {Abdo}},
  \bibinfo {author} {\bibfnamefont {Markus}\ \bibnamefont {Brink}}, \bibinfo
  {author} {\bibfnamefont {Andrew}\ \bibnamefont {Cross}}, \bibinfo {author}
  {\bibfnamefont {Jerry~M.}\ \bibnamefont {Chow}}, \ and\ \bibinfo {author}
  {\bibfnamefont {Jay~M.}\ \bibnamefont {Gambetta}},\ }\bibfield  {title}
  {\enquote {\bibinfo {title} {Demonstration of weight-four parity measurements
  in the surface code architecture},}\ }\href {\doibase
  10.1103/physrevlett.117.210505} {\bibfield  {journal} {\bibinfo  {journal}
  {Phys. Rev. Lett.}\ }\textbf {\bibinfo {volume} {117}},\ \bibinfo {pages}
  {210505} (\bibinfo {year} {2016})}\BibitemShut {NoStop}%
\bibitem [{\citenamefont {Bacon}(2006)}]{Bacon2006}%
  \BibitemOpen
  \bibfield  {author} {\bibinfo {author} {\bibfnamefont {Dave}\ \bibnamefont
  {Bacon}},\ }\bibfield  {title} {\enquote {\bibinfo {title} {Operator quantum
  error-correcting subsystems for self-correcting quantum memories},}\ }\href
  {\doibase 10.1103/PhysRevA.73.012340} {\bibfield  {journal} {\bibinfo
  {journal} {Phys. Rev. A}\ }\textbf {\bibinfo {volume} {73}},\ \bibinfo
  {pages} {012340} (\bibinfo {year} {2006})}\BibitemShut {NoStop}%
\bibitem [{\citenamefont {Bombin}\ and\ \citenamefont
  {Martin-Delgado}(2006)}]{Bombin2006}%
  \BibitemOpen
  \bibfield  {author} {\bibinfo {author} {\bibfnamefont {H.}~\bibnamefont
  {Bombin}}\ and\ \bibinfo {author} {\bibfnamefont {M.~A.}\ \bibnamefont
  {Martin-Delgado}},\ }\bibfield  {title} {\enquote {\bibinfo {title}
  {Topological quantum distillation},}\ }\href {\doibase
  10.1103/PhysRevLett.97.180501} {\bibfield  {journal} {\bibinfo  {journal}
  {Phys. Rev. Lett.}\ }\textbf {\bibinfo {volume} {97}},\ \bibinfo {pages}
  {180501} (\bibinfo {year} {2006})}\BibitemShut {NoStop}%
\bibitem [{\citenamefont {Chamberland}\ \emph {et~al.}(2019)\citenamefont
  {Chamberland}, \citenamefont {Zhu}, \citenamefont {Yoder}, \citenamefont
  {Hertzberg},\ and\ \citenamefont {Cross}}]{Chamberland2019}%
  \BibitemOpen
  \bibfield  {author} {\bibinfo {author} {\bibfnamefont {C.}~\bibnamefont
  {Chamberland}}, \bibinfo {author} {\bibfnamefont {G.}~\bibnamefont {Zhu}},
  \bibinfo {author} {\bibfnamefont {T.~J.}\ \bibnamefont {Yoder}}, \bibinfo
  {author} {\bibfnamefont {J.~B.}\ \bibnamefont {Hertzberg}}, \ and\ \bibinfo
  {author} {\bibfnamefont {A.~W.}\ \bibnamefont {Cross}},\ }\bibfield  {title}
  {\enquote {\bibinfo {title} {Topological and subsystem codes on low-degree
  graphs with flag qubits},}\ }\href {https://arxiv.org/abs/1907.09528}
  {\bibfield  {journal} {\bibinfo  {journal} {arXiv:1907.09528}\ } (\bibinfo
  {year} {2019})}\BibitemShut {NoStop}%
\bibitem [{\citenamefont {Li}\ \emph {et~al.}(2019)\citenamefont {Li},
  \citenamefont {Miller}, \citenamefont {Newman}, \citenamefont {Wu},\ and\
  \citenamefont {Brown}}]{Li2019s}%
  \BibitemOpen
  \bibfield  {author} {\bibinfo {author} {\bibfnamefont {Muyuan}\ \bibnamefont
  {Li}}, \bibinfo {author} {\bibfnamefont {Daniel}\ \bibnamefont {Miller}},
  \bibinfo {author} {\bibfnamefont {Michael}\ \bibnamefont {Newman}}, \bibinfo
  {author} {\bibfnamefont {Yukai}\ \bibnamefont {Wu}}, \ and\ \bibinfo {author}
  {\bibfnamefont {Kenneth~R.}\ \bibnamefont {Brown}},\ }\bibfield  {title}
  {\enquote {\bibinfo {title} {2d compass codes},}\ }\href {\doibase
  10.1103/PhysRevX.9.021041} {\bibfield  {journal} {\bibinfo  {journal} {Phys.
  Rev. X}\ }\textbf {\bibinfo {volume} {9}},\ \bibinfo {pages} {021041}
  (\bibinfo {year} {2019})}\BibitemShut {NoStop}%
\bibitem [{\citenamefont {Wood}\ and\ \citenamefont
  {Gambetta}(2018)}]{Wood2017}%
  \BibitemOpen
  \bibfield  {author} {\bibinfo {author} {\bibfnamefont {Christopher~J.}\
  \bibnamefont {Wood}}\ and\ \bibinfo {author} {\bibfnamefont {Jay~M.}\
  \bibnamefont {Gambetta}},\ }\bibfield  {title} {\enquote {\bibinfo {title}
  {Quantification and characterization of leakage errors},}\ }\href {\doibase
  10.1103/PhysRevA.97.032306} {\bibfield  {journal} {\bibinfo  {journal} {Phys.
  Rev. A}\ }\textbf {\bibinfo {volume} {97}},\ \bibinfo {pages} {032306}
  (\bibinfo {year} {2018})}\BibitemShut {NoStop}%
\bibitem [{\citenamefont {Chen}\ \emph {et~al.}(2016)\citenamefont {Chen},
  \citenamefont {Kelly}, \citenamefont {Quintana}, \citenamefont {Barends},
  \citenamefont {Campbell}, \citenamefont {Chen}, \citenamefont {Chiaro},
  \citenamefont {Dunsworth}, \citenamefont {Fowler}, \citenamefont {Lucero},
  \citenamefont {Jeffrey}, \citenamefont {Megrant}, \citenamefont {Mutus},
  \citenamefont {Neeley}, \citenamefont {Neill}, \citenamefont {O'Malley},
  \citenamefont {Roushan}, \citenamefont {Sank}, \citenamefont {Vainsencher},
  \citenamefont {Wenner}, \citenamefont {White}, \citenamefont {Korotkov},\
  and\ \citenamefont {Martinis}}]{Chen2016}%
  \BibitemOpen
  \bibfield  {author} {\bibinfo {author} {\bibfnamefont {Zijun}\ \bibnamefont
  {Chen}}, \bibinfo {author} {\bibfnamefont {Julian}\ \bibnamefont {Kelly}},
  \bibinfo {author} {\bibfnamefont {Chris}\ \bibnamefont {Quintana}}, \bibinfo
  {author} {\bibfnamefont {R.}~\bibnamefont {Barends}}, \bibinfo {author}
  {\bibfnamefont {B.}~\bibnamefont {Campbell}}, \bibinfo {author}
  {\bibfnamefont {Yu}~\bibnamefont {Chen}}, \bibinfo {author} {\bibfnamefont
  {B.}~\bibnamefont {Chiaro}}, \bibinfo {author} {\bibfnamefont
  {A.}~\bibnamefont {Dunsworth}}, \bibinfo {author} {\bibfnamefont {A.~G.}\
  \bibnamefont {Fowler}}, \bibinfo {author} {\bibfnamefont {E.}~\bibnamefont
  {Lucero}}, \bibinfo {author} {\bibfnamefont {E.}~\bibnamefont {Jeffrey}},
  \bibinfo {author} {\bibfnamefont {A.}~\bibnamefont {Megrant}}, \bibinfo
  {author} {\bibfnamefont {J.}~\bibnamefont {Mutus}}, \bibinfo {author}
  {\bibfnamefont {M.}~\bibnamefont {Neeley}}, \bibinfo {author} {\bibfnamefont
  {C.}~\bibnamefont {Neill}}, \bibinfo {author} {\bibfnamefont {P.~J.~J.}\
  \bibnamefont {O'Malley}}, \bibinfo {author} {\bibfnamefont {P.}~\bibnamefont
  {Roushan}}, \bibinfo {author} {\bibfnamefont {D.}~\bibnamefont {Sank}},
  \bibinfo {author} {\bibfnamefont {A.}~\bibnamefont {Vainsencher}}, \bibinfo
  {author} {\bibfnamefont {J.}~\bibnamefont {Wenner}}, \bibinfo {author}
  {\bibfnamefont {T.~C.}\ \bibnamefont {White}}, \bibinfo {author}
  {\bibfnamefont {A.~N.}\ \bibnamefont {Korotkov}}, \ and\ \bibinfo {author}
  {\bibfnamefont {John~M.}\ \bibnamefont {Martinis}},\ }\bibfield  {title}
  {\enquote {\bibinfo {title} {Measuring and suppressing quantum state leakage
  in a superconducting qubit},}\ }\href {\doibase
  10.1103/PhysRevLett.116.020501} {\bibfield  {journal} {\bibinfo  {journal}
  {Phys. Rev. Lett.}\ }\textbf {\bibinfo {volume} {116}},\ \bibinfo {pages}
  {020501} (\bibinfo {year} {2016})}\BibitemShut {NoStop}%
\bibitem [{\citenamefont {Macklin}\ \emph {et~al.}(2015)\citenamefont
  {Macklin}, \citenamefont {O'Brien}, \citenamefont {Hover}, \citenamefont
  {Schwartz}, \citenamefont {Bolkhovsky}, \citenamefont {Zhang}, \citenamefont
  {Oliver},\ and\ \citenamefont {Siddiqi}}]{Macklin2015}%
  \BibitemOpen
  \bibfield  {author} {\bibinfo {author} {\bibfnamefont {C.}~\bibnamefont
  {Macklin}}, \bibinfo {author} {\bibfnamefont {K.}~\bibnamefont {O'Brien}},
  \bibinfo {author} {\bibfnamefont {D.}~\bibnamefont {Hover}}, \bibinfo
  {author} {\bibfnamefont {M.~E.}\ \bibnamefont {Schwartz}}, \bibinfo {author}
  {\bibfnamefont {V.}~\bibnamefont {Bolkhovsky}}, \bibinfo {author}
  {\bibfnamefont {X.}~\bibnamefont {Zhang}}, \bibinfo {author} {\bibfnamefont
  {W.~D.}\ \bibnamefont {Oliver}}, \ and\ \bibinfo {author} {\bibfnamefont
  {I.}~\bibnamefont {Siddiqi}},\ }\bibfield  {title} {\enquote {\bibinfo
  {title} {A near-quantum-limited {Josephson} traveling-wave parametric
  amplifier},}\ }\href {\doibase 10.1126/science.aaa8525} {\bibfield  {journal}
  {\bibinfo  {journal} {Science}\ }\textbf {\bibinfo {volume} {350}},\ \bibinfo
  {pages} {307--310} (\bibinfo {year} {2015})}\BibitemShut {NoStop}%
\bibitem [{\citenamefont {Johansson}\ \emph {et~al.}(2013)\citenamefont
  {Johansson}, \citenamefont {Nation},\ and\ \citenamefont
  {Nori}}]{Johansson2013a}%
  \BibitemOpen
  \bibfield  {author} {\bibinfo {author} {\bibfnamefont {J.~R.}\ \bibnamefont
  {Johansson}}, \bibinfo {author} {\bibfnamefont {P.~D.}\ \bibnamefont
  {Nation}}, \ and\ \bibinfo {author} {\bibfnamefont {Franco}\ \bibnamefont
  {Nori}},\ }\bibfield  {title} {\enquote {\bibinfo {title} {{QuTiP} 2: {A}
  {Python} framework for the dynamics of open quantum systems},}\ }\href
  {\doibase 10.1016/j.cpc.2012.11.019} {\bibfield  {journal} {\bibinfo
  {journal} {Comput. Phys. Commun.}\ }\textbf {\bibinfo {volume} {184}},\
  \bibinfo {pages} {1234--1240} (\bibinfo {year} {2013})}\BibitemShut {NoStop}%
\bibitem [{\citenamefont {Wiseman}\ and\ \citenamefont
  {Milburn}(2010)}]{Wiseman2010}%
  \BibitemOpen
  \bibfield  {author} {\bibinfo {author} {\bibfnamefont {H.}~\bibnamefont
  {Wiseman}}\ and\ \bibinfo {author} {\bibfnamefont {G.}~\bibnamefont
  {Milburn}},\ }\href@noop {} {\emph {\bibinfo {title} {Quantum Measurement and
  Control}}}\ (\bibinfo  {publisher} {Cambridge University Press},\ \bibinfo
  {year} {2010})\BibitemShut {NoStop}%
\end{thebibliography}%

\end{document}